\documentclass[twocolumn,longtable,tighten]{aastex63}
\usepackage{apjfonts}
\usepackage{amsbsy}
\usepackage{epstopdf}
\usepackage{graphicx}
\newcommand{\hour}{\mbox{$^{\rm h}$}}

\newcommand{\minute}{\mbox{$^{\rm m}$}}

\DeclareMathAlphabet{\mathsfsl}{OT1}{cmss}{m}{sl}
\shorttitle{Probing Interstellar Grain Growth through Polarimetry}
\shortauthors{Vaillancourt et al.}

\begin{document}

\title{Probing Interstellar Grain Growth Through Polarimetry in the Taurus Cloud Complex}

\author[0000-0001-8916-1828]{John E. Vaillancourt}
\affiliation{Lincoln Laboratory, Massachusetts Institute of Technology, 244 Wood St., Lexington, MA 02421-6426}
\affiliation{SOFIA Science Center, Universities Space Research Association,
  NASA Ames Research Center, Moffett Field, CA 94035}

\author[0000-0001-6717-0686]{B-G Andersson}
\affiliation{SOFIA Science Center, Universities Space Research Association,
  NASA Ames Research Center, Moffett Field, CA 94035}

\author[0000-0002-9947-4956]{Dan P. Clemens} 
\affiliation{Institute for Astrophysical Research and Department of
    Astronomy, Boston University, 725 Commonwealth Ave., Boston, MA
    02215}

\author{Vilppu Piirola}
\affiliation{Tuorla Observatory, University of Turku, FI-21500
  Piikki\"{o}, Finland}

\author[0000-0003-2017-0982]{Thiem Hoang}
\affiliation{Korean Astronomy and Space Science Institute, Daejeon
  34055, Korea, and Korean University of Science and Technology,
  Daejeon, 34113, Korea}

\author{Eric E. Becklin}
\affiliation{SOFIA Science Center, Universities Space Research Association,
  NASA Ames Research Center, Moffett Field, CA 94035}
\affiliation{Department of Physics \& Astronomy, University of California, Los Angeles}

\author[0000-0002-2957-3924]{Miranda Caputo} 
\affiliation{SOFIA Science Center, Universities Space Research
  Association, NASA Ames Research Center, Moffett Field, CA 94035}
\affiliation{Ritter Astrophysical Research Center, University of
  Toledo Dept. of Physics and Astronomy, 2801 W. Bancroft St. Toledo,
  Ohio 43606}

\begin{abstract}
  The optical and near-infrared (OIR) polarization of starlight is typically understood to arise from the dichroic extinction of that light by dust grains whose axes are aligned with respect to a local magnetic field. The size distribution of the aligned grain population can be constrained by measurements of the wavelength dependence of the polarization.  The leading physical model for producing the alignment is that of radiative alignment torques (RAT), which predicts that the most efficiently aligned grains are those with sizes larger than the wavelengths of light composing the local radiation field. Therefore, for a given grain size distribution, the wavelength at which the polarization reaches a maximum ($\lambda_\mathrm{max}$) should correlate with the characteristic reddening along the line of sight between the dust grains and the illumination source.  A correlation between $\lambda_\mathrm{max}$ and reddening \deleted{with respect to the observer} has been previously established for extinctions up to $A_V\approx4$\,mag.  We extend the study of this relationship to a \deleted{significantly} larger sample of stars in the Taurus cloud complex, including extinctions \deleted{higher than} $A_V>10$\,mag. We confirm the earlier results for $A_V<4$\,mag, but find that the $\lambda_\mathrm{max}$ vs.\ $A_V$ relationship bifurcates above $A_V\approx4$\,mag, with part of the sample continuing the previously observed \deleted{linear} relationship.  The remaining sample exhibits a \deleted{significantly} steeper rise in $\lambda_\mathrm{max}$ vs.\ $A_V$.  We propose that the data exhibiting the steep rise represent lines of sight of \deleted{localized} high-density ``clumps'', where grain coagulation has taken place.  We present RAT-based modeling supporting these hypotheses.  These results indicate that multi-band OIR polarimetry is a powerful tool for tracing grain growth in molecular clouds, independent of uncertainties in the dust temperature and emissivity.\deleted{ distributions.}
  \end{abstract}


\section{Introduction} \label{sec-intro}

Starlight passing through the interstellar medium (ISM) is typically
polarized at the level of a few percent. The upper envelope of the
polarization fraction correlates well with the extinction
\citep[e.g.,][]{hiltner1949a,serkowski1968,fosalba2002}, and the
position angle is in good agreement with independent measurements of
the interstellar magnetic-field orientation
\citep[e.g.,][]{spoelstra1984,scarrott1987,page2007}. Thus, the
observed optical and near-infrared (OIR) polarization, and the
complementary far-infrared polarized emission
\citep[e.g.,][]{cudlip1982,dotson2000}, have long been attributed to
asymmetric dust grains aligned with the interstellar magnetic field
\citep[e.g.,][]{hiltner1949b,hildebrand1988}. Polarimetry can provide
a powerful tool for probing interstellar magnetic fields
\citep{davis1951b,chandrasekhar1953} if the process of dust grain
alignment
can be understood.

Attempts to explain grain alignment began with its discovery
\citep{hiltner1949a,hall1949} and included ferromagnetic alignment
\citep{spitzer1949}, mechanical alignment \citep{gold1952}, and
paramagnetic relaxation (\citealt{davis1951a}).  Significant progress
has been achieved in the last decade, both theoretically and
observationally \citep[see reviews
by][]{lazarian2015,bga2015a,bga2015b}.  New calculations
\citep{lazarian1999b,hoang2016} and observations \citep{hough2008}
have shown that paramagnetic relaxation (or its
modifications---\citealt{purcell1976, mathis1986}) likely cannot
provide an explanation of the observed interstellar polarization.  In
contrast, a quantitative theory based on direct radiative alignment
torques (RATs; e.g., \citealt{draine1996,draine1997}) now provides
specific testable predictions \citep{lazarian2007,hoang2008}.  The aim
of this work is to extend tests of the prediction that the alignment
efficiency depends on the color of the local radiation field and the
size distribution of the grains.

The basic requirement for grain alignment in RAT theory is that a
grain of effective radius $a$, with net helicity, be exposed to an
anisotropic radiation field with a wavelength $\lambda$ that is less
than the grain diameter \citep{lazarian2007}.  Grain helicity is
satisfied for any irregularly shaped grain. Radiation fields in the
ISM are almost always anisotropic since the grain is either located
close to a discrete radiation source (star) and/or the interstellar
cloud in which it is embedded has a well defined density gradient and
hence a net radiation-flow vector.  The size constraint follows in a
manner similar to Mie scattering theory (e.g., \citealt{martin1974})
such that the coupling of the aligning radiation to the grains drops
rapidly for $\lambda>2a$ \citep{lazarian2007}\footnote{The radiative
  torque efficiency does not fully disappear at $\lambda > 2a$, but
  drops as ($\lambda/a)^{-\alpha}$, with $\alpha\sim3$--4.}.  From
these requirements, it follows that the color and intensity of the
radiation field are key factors in radiative grain alignment.  Support
of this dependence comes from observations of correlations between
radiation field strengths and polarization
\citep{whittet2008,medan2019}

Tests of the RAT predictions require measurements that trace grain
alignment efficiency, preferably as a function of grain size.  The
fractional polarization $p/\tau$ (where $\tau$ is the optical depth at
some chosen wavelength) nominally traces the alignment
\citep{whittet2008}, but is not directly sensitive to changes in the
grain size distribution and cannot be used to distinguish
between line of sight alignment variations and changes in magnetic
field structure and turbulence \citep{jones1989}.

A better measure of the size distribution of aligned grains is the
wavelength dependence of the polarization because relative variations
in this spectrum are independent of the grain alignment orientation
with respect to the line of sight.  Empirically, the wavelength
dependence of interstellar polarization is parameterized by the
wavelength at which the polarization peaks, $\lambda_\mathrm{max}$,
using the function
\begin{equation} 
  \frac{p(\lambda)}{p_\mathrm{max}} = \exp \left[ -K\; \ln^2 \left( \frac{\lambda}{\lambda_\mathrm{max}}\right) \right],
  \label{serkowski} 
\end{equation}
referred to as the ``Serkowski relation'' \citep{serkowski1975} if the
parameter $K$ is set to the fixed value 1.15, or the ``Wilking
relation'' \citep{wilking1980} if $K$ is used as a fitting parameter.
\deleted{\citet{codina1976} first suggested that $K$ varies in the ISM
  and \citet{wilking1982} and \citet{whittet1992} showed that $K$ and
  $\lambda_\mathrm{max}$ are correlated.  As $K$ parametrizes the
  width of the curve, a dependence on $\lambda_\mathrm{max}$ is
  expected under RAT alignment, since---for a fixed maximum grain
  size---shifting the minimum size of aligned grains will shift both
  the average grain size (and $\lambda_\mathrm{max}$), but will also
  narrow the size distribution.}

\citet{kim1995} showed theoretically that $\lambda_\mathrm{max}$
depends on the average size of aligned grains, and is especially
sensitive to the size of the smallest grains in the distribution.  By
inverting the observed extinction and polarization curves, they
showed, in agreement with earlier studies \citep{mathis1977}, that the
overall dust size distribution extends to very small grains
($a \sim 0.01\,\micron$), but that the aligned silicate grains in the
diffuse ISM typically do not have sizes smaller than
$a=0.04$--$0.05\,\micron$.  \deleted{Given the RAT alignment criterion
  ($\lambda<2a$) this size limit may be due to the lack of
  interstellar radiation shortward of the Lyman limit
  (0.0912\,\micron).}  Since interstellar extinction increases as the
wavelength of incident radiation becomes bluer, RAT alignment predicts
that the size of the smallest aligned grain will increase as the
radiation is reddened into the cloud---hence $\lambda_\mathrm{max}$
should be correlated with the extinction $A_V$.

Using multi-band polarimetry, \citet{whittet2001} noted a weak
correlation between $\lambda_\mathrm{max}$ and $A_V$ in their Taurus
sample.  \citet{bga2007} re-analyzed those data and showed that, if
several observational biases are taken into account (most importantly
line of sight where the visual extinction and the FIR color
temperatures were inconsistent), this correlation becomes well defined
and is present for all six nearby interstellar clouds probed in their
study.  However, the observational samples analyzed in \citet{bga2007}
were of limited size and only covered the extinction range up to
$A_V\sim 4$\,mag, with the exception of a very small number of higher
extinction data points.  For the Taurus cloud, \citet{bga2007} found
\begin{equation}
  \lambda_\mathrm{max}= (0.53\pm0.01\;\micron) \, + \, (0.020\pm0.004\;
  \micron\,\mathrm{mag}^{-1})\, A_V.
  \label{whittet_fit}
\end{equation}
If the RAT prediction is correct (that the minimum aligned-grain size
will increase as the radiation is reddened), and the grain size
distribution is constant, then this $\lambda_\mathrm{max}$-$A_V$
relation should continue to higher levels of extinction, where very
few targets have been previously observed. To date, the location of
the peak polarization towards one star, Elias 3-16 at
$A_V=24.1\pm0.1$\,mag \citep{murakawa2000} and
$\lambda_\mathrm{max}=1.08\pm0.08\;\micron$ \citep{hough1988}, is
consistent with the linear relation in equation~(\ref{whittet_fit}).

Our goal here is to test the low-extinction results found by
\citet{whittet2001} and \citet{bga2007} and determine if the
$\lambda_\mathrm{max}$-$A_V$ trend continues at $A_V>4$\,mag.  We have
chosen to focus on the Taurus cloud complex as it is well studied in
many tracers and covers a wide range of extinctions (see
\citet{kenyon2008} for a review).  Of particular importance for our
purposes, Taurus, being near-by, has background field-stars that tend
to be reasonably bright and therefore amenable to high quality
polarimetry.  Several authors have surveyed the cloud in polarimetry
\citep[][and refs therein]{whittet1992,whittet2008} and near infrared
photometry \citep[e.g.][]{shenoy2008}

We present two additional data sets for stars in the Taurus cloud
complex not previously observed in polarization at multiple
wavelengths. Our target stars have been selected to be true background
stars (behind the cloud complex from our LOS), thereby avoiding the
complications of grain alignment variations induced in the
circumstellar environment of embedded stars \citep{whittet2008}.
These new observations are described in Section~\ref{sec-obs}.
Section~\ref{sec-results} presents fits to the Serkowski and Wilking
relations for these new data sets and includes discussion of the
results of some individual stars. Within Sections
\ref{sec-obs}--\ref{sec-results} there are three tables for each of
the two data sets: 1) Tables \ref{turpol_source} and \ref{sourcetab}
contain lists of the observed stars along with characteristics such as
apparent brightness, extinction, and distance; 2) Tables
\ref{tbl-results-turpol} and \ref{tbl-results} contain the Stokes
parameters fitted to the polarization data for each star and each
bandpass; and 3) Tables \ref{tbl-serkowskifits-turpol} and
\ref{tbl-serkowskifits} contain fits to the Serkowski and Wilking
relations for each star.  Section~\ref{sec-discussion} examines the
$\lambda_\mathrm{max}$-$A_V$ relation and its variations and uses an
RAT grain-alignment model to interpret the observational results.  Our
results and conclusions are summarized in
Section~\ref{Conclusion_sec}.

\begin{figure*}
  \plotone{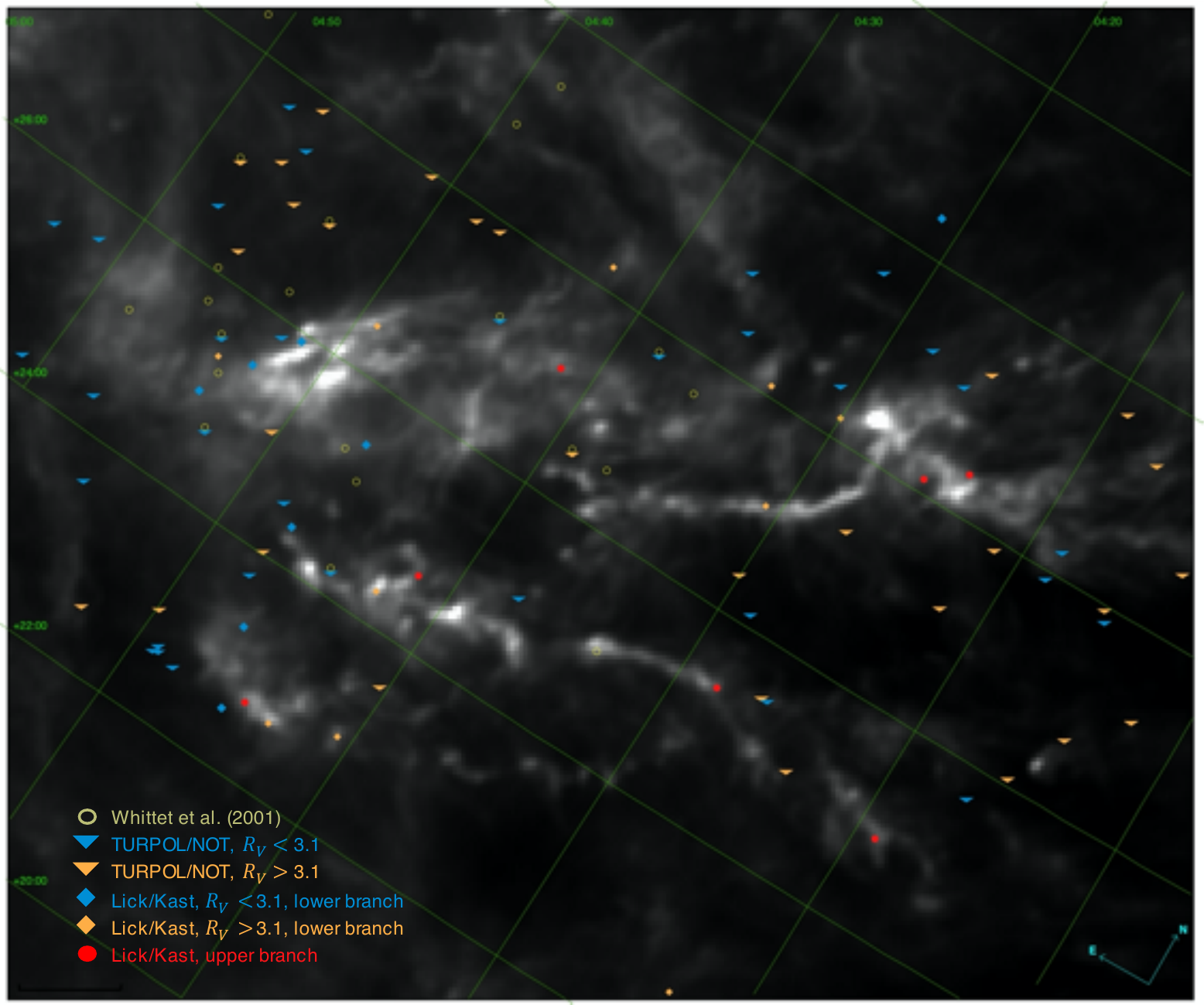}
  \caption{Location of the stars observed in polarization in the
    Taurus Cloud Complex, plotted on the \emph{Planck} $350\,\micron$
    map \citep{Planck2018iii}.  Stars in the sample of
    \citet{whittet2001} are shown as yellow circles. Stars observed by
    NOT/TURPOL are shown as triangles (blue and orange) and stars
    observed with Lick/Kast on the ``lower branch'' of the
    $\lambda_\mathrm{max}$-$A_V$ relation
    (Section~\ref{sec-discussion1}) are shown as diamonds (blue and
    orange).  In both data sets, stars with $R_V<3.1$ are blue and
    stars with $R_V>3.1$ are orange. Red circles are Lick/Kast targets
    on the ``upper branch'' of the $\lambda_\mathrm{max}$-$A_V$
    relation (Section~\ref{sec-discussion1}). }
  \label{fig-map}
\end{figure*}

\begin{figure}
  \plotone{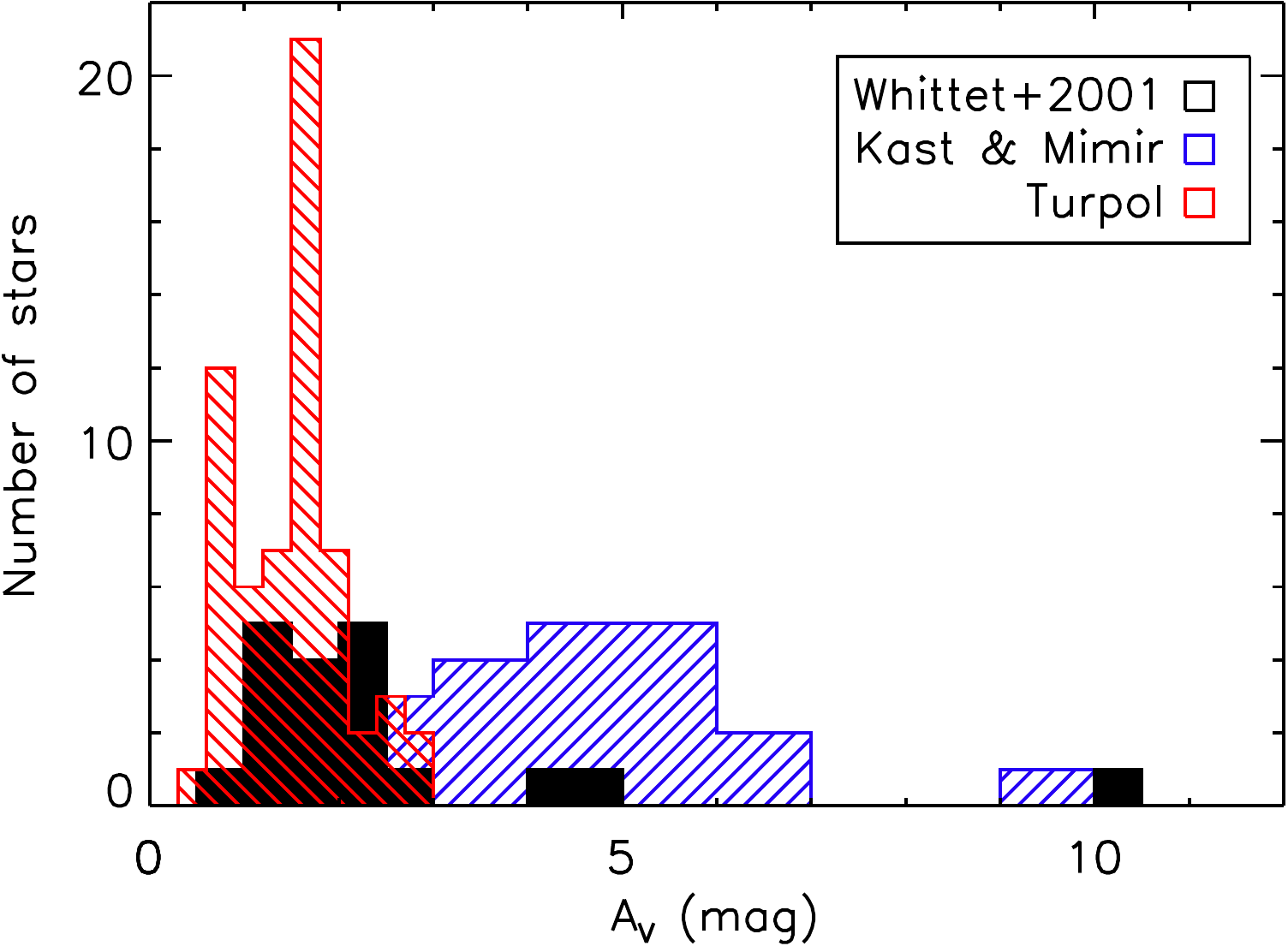}
  \caption{Distribution of extinctions. The number of stars in several
    extinction bins for the work of \citet{whittet2001}, as well as
    the data presented in this work obtained using the Kast
    (Section~\ref{sec-kast}), Mimir (Section~\ref{sec-mimir}), and
    Turpol (Section~\ref{sec-turpol}) instruments. The Turpol data
    cover an extinction range similar to that of the
    \citeauthor{whittet2001}\ data, while the Kast and Mimir data
    extend the data sets to higher extinction levels.}
  \label{fig-hist}
\end{figure}

\section{Observations \& Data Reduction} \label{sec-obs}

\subsection{Target Selection}

Our sample of low-extinction stars (Table \ref{turpol_source}) was
based on the \citet{wright2003} survey of spectroscopically-classified
Tycho targets \citep{tycho2000}, screening out sightlines with
anomalous visual extinctions, following the procedure in
\citet{bga2007}.  As described in that paper, such anomalous lines of
sight represent material where, based on comparisons of $A_V$ and FIR
dust color-temperature, the line-of-sight visual extinction ($A_V$)
does not accurately represent the effective radiation field seen by
the dust.  This can be due to the geometry of the cloud or the
presence of near-by discrete radiation sources.  We selected stars
from two fields of 3 degree radius centered at Right Ascension,
Declination (J2000): ($4\hour15\minute$, $28\arcdeg0\arcmin$) and
($4\hour40\minute$, $25\arcdeg30\arcmin$).  Using Tycho and 2MASS
\citep{skrutskie2006} photometry, we derived visual extinctions
($A_V$) and the total-to-selective-extinction ratio ($R_V$).
 
The high-extinction stellar sample list was generated using stars from
the surveys by \citet{teixeira1999}, \citet{murakawa2000}, and
\citet[][hereafter SWIW]{shenoy2008} (Table \ref{sourcetab}).  We
screened these for embedded sources by rejecting any targets that show
associated flux in the 60 or $100\,\micron$ bands of IRAS\@.  At the
time of the observations,
NOMAD\footnote{\url{http://www.usno.navy.mil/USNO/astrometry/optical-IR-prod/nomad}}
photometry was available for most stars at $B$-, $V$-, and $R$-bands,
as is 2MASS photometry at $J$-, $H$-, and $K_s$-bands. Because of the
marginal photometric accuracy of the NOMAD data, the visual
extinctions are, where available, taken from SWIW, and are based on
the relation $A_V = (5.3\pm0.3) \times E_{J-K}$ (c.f., SWIW\@).  The
selected sample contains the 25 stars listed in Table~\ref{sourcetab}
and spans extinctions from $A_V=3$ to 10 magnitudes
(Figure~\ref{fig-hist}).

In order to sample the material in the cloud we also limit the stellar
sample to those stars behind the cloud. Distances from the Gaia DR2
parallax survey \citep{gaiadr2} are shown in Tables
\ref{turpol_source} and \ref{sourcetab}, and discussed further in
Section~\ref{sec-stars}.  Figure~\ref{fig-map} shows the location of
all stars in our sample within the structure of the Taurus Cloud
Complex, which is traced by the total dust emission at
$160\,\micron$. (The stars are plotted as different colors based on
$R_V$ and the $\lambda_\mathrm{max}$-$A_V$ relation discussed in
Section~\ref{sec-discussion1}.)

\subsection{Optical Photo-polarimetry} \label{sec-turpol}

We used the TurPol instrument on the 2.5\,m Nordic Optical Telescope
(NOT) at Roque de los Muchachos Observatory on La Palma, during the
nights of 2007 November 3--6, to perform broad-band multicolor
polarimetry of stars background to the Taurus molecular cloud. The
instrument provides simultaneous measurements in five passbands close
to the $U\! BV\! RI$ system, by using four dichroic filters to split
the light into the different spectral regions \citep{Piirola1988}.
The instrument was used in linear polarimetry mode. The half-wave
plate (HWP) was rotated in 22.5 degree steps, with one complete
observation of linear polarization consisting of eight integrations.
Standard stars were observed each night to calibrate the position
angle zero-point and determine the instrumental polarization.

\subsection{Optical Spectropolarimetry} \label{sec-kast}

We performed spectropolarimetric observations of high-extinction lines
of sight using the red-channel (0.4--$1.1\,\micron$) of the Kast
spectropolarimeter \citep*{miller1988} on the 3\,m Shane telescope of
Lick Observatory during the nights of 2009 November 15--17 (UT\@).
The incident polarization was modulated by rotating a HWP through
eight 22.5 degree steps, spanning the range 0--157.5 degrees, and
integrating for 20--60 seconds per step, depending on the stellar
brightness.  Spectra in both orthogonal modes of linear polarization
were simultaneously imaged onto a $1200\times400$\,pixel CCD with
$27\,\micron$ pixels (manufactured by EG\&G Reticon;
\citealt{cizdziel1990}).  Typical observations used wide slits
($3\arcsec$--$5\arcsec$) yielding stellar images of $\sim$4 camera
pixels full-width at half maxium (FWHM\@). In conjuntion with a 300
lines per millimeter grating, at a blaze angle corresponding to
7500\,\AA, the resulting spectra have resolutions of
$\approx$20\,\AA\@ (FWHM).

Detailed discussion of the stellar polarization data analyses are
presented in Appendix~\ref{sec-panalysis} along with calibration
results and procedures in Appendix~\ref{sec-pcal}.  Wavelength
calibration was achieved by regularly observing arc lamps composed of
argon, helium, and neon.  Flat-field corrections utilized dome flats,
separately normalizing the two orthogonal polarization components in
each image and for each of the eight HWP angles.  Two unpolarized
standard stars were used to measure the level of instrument
polarization and four high-polarization standard stars to calibrate
the polarization position angle.  Additionally, the unpolarized
standards were used in conjunction with a polarizing filter in order
to characterize the wavelength dependence of the polarization
efficiency and offset angles of the HWP.

Spectral classification of the spectro-polarimetry sample stars was
accomplished by comparing the Stokes $I$ spectra with the standard
sequence from \citet{jacoby1984} and other classification estimates in
the literature \citep[e.g.][]{murakawa2000}. The spectral
classification was performed independently by two astronomers whose
separate spectral class estimates agreed to within 1--2
subclasses. Given the limited spectral resolution (20\,\AA), reliable
luminosity class determinations are not possible and we have,
therefore, assigned luminosity class V (Main Sequence) to all stars
earlier than G0 and class III (Giant) to later spectral types. Using
parallaxes, archival photometry, and absolute magnitudes measured by
Gaia \citep{gaiadr2} we found that this assumption held true for all
stars in our sample except for SWIW\,040, SWIW\,148, and Tamura 17.

\subsection{Near-Infrared Polarimetry} \label{sec-mimir}

Imaging polarimetric observations in the near-infrared (NIR) $H$-band
($1.63\,\micron$) took place on the nights of 2011 September 18 and 19
and again on 2012 January 11 using the Mimir instrument
\citep{clemens2007} on the 1.83\,m Perkins telescope, located outside
Flagstaff, AZ\@. Mimir used a rotating compound HWP, an MKO $H$-band
filter, a PK50 long-wave blocker, and a fixed wire-grid to perform
step-and-integrate polarimetry over a $10\arcmin \times 10\arcmin$
field of view at 0.6~arcsec per pixel onto a $1024^2$\,pixel Aladdin
III InSb detector array. A single observation sequence consisted of
pointing the telescope to each of six sky dither positions, located
about a 15\,arcsec wide hexagon, and collecting 16 images, each
through HWP orientations offset by $22.5\arcdeg$, to sample one
complete turn of the HWP\@. This yielded four independent sets of
Stokes $q$ and $u$. In-dome calibrations included linearity,
polarization flat-fields, and dark current images. Sky calibrations
consisted of observations of fields containing multiple polarization
standard stars. Further detail of the data collection steps, data
processing and polarimetric extractions are to be found in
\citet{clemens2012b,clemens2012a}.

Some of the stars in Table~\ref{sourcetab} are the brightest observed
for polarimetry by Mimir and some of the 2011 observations had poorly
matched exposure times. The 2012 observations used reduced integration
times (0.25--1.75~seconds vs.\ 0.65--3.25~seconds per exposure),
yielding lower uncertainties. For the non-saturated stars observed in
both runs, variance-weighted averaging of Stokes $q$ and $u$ was
performed and propagated into polarization percentages and equatorial
position angles and their associated uncertainties.  Single
observations of bright stars are limited to about 0.15\%--0.2\%
polarization percentage uncertainty and about $1\arcdeg$--$2\arcdeg$
position angle uncertainty.

%
%
\startlongtable
\begin{deluxetable*}{lccDcccc}
\tablecaption{TurPol Sources \label{turpol_source}}             
\tablehead{\colhead{} & \colhead{RA (J2000)} & \colhead{Dec.\
    (J2000)} & \multicolumn2c{$V$\tablenotemark{a}} & \colhead{$R_V$} & \colhead{$\sigma_{R_V}$} & \colhead{$A_V$} & \colhead{$\sigma_{A_V}$} \\
  \colhead{Star} & \colhead{(hh:mm:ss.s)} & \colhead{(dd:mm:ss.s)} & \multicolumn2c{(mag.)} & \colhead{} & \colhead{} & \colhead{(mag.)} & \colhead{(mag.)} }

\decimals
\startdata
PPM\,93181  & 4:02:15.2 & 27:10:36.7 &  9.8  & 6.24 & 2.61  &  0.74 & 0.08 \\
PPM\,93195  & 4:03:38.8 & 28:39:42.3 &   8.1 & 4.10 & 0.64  &  1.09 & 0.09 \\
PPM\,93213  & 4:04:22.4 & 26:42:12.8 &  10.8 & 6.41 & 5.33  &  0.85 & 0.14 \\
PPM\,93236  & 4:05:28.8 & 27:52:46.7 &   8.4 & 2.24 & 0.32  &  0.80 & 0.08 \\
PPM\,92238  & 4:05:36.1 & 26:05:18.8 &  10.5 & 4.37 & 1.01  &  0.88 & 0.08 \\
PPM\,93241  & 4:05:44.7 & 27:58:47.7 &  10.6 & 3.30 & 1.67  &  0.60 & 0.12 \\
PPM\,93260  & 4:06:41.4 & 25:43:19.7 &   8.0 & 2.94 & 0.67  &  0.58 & 0.08 \\
PPM\,93265  & 4:07:05.1 & 29:26:25.9 &  10.0 & 4.87 & 0.56  &  2.71 & 0.09 \\
PPM\,93280  & 4:08:40.8 & 28:14:56.4 &  10.6 & 2.57 & 0.39  &  2.23 & 0.11 \\
PPM\,93281  & 4:08:41.3 & 27:56:40.0 &  10.8 & 2.64 & 0.79  &  1.02 & 0.11 \\
PPM\,93289  & 4:09:21.1 & 29:43:47.4 &   9.9 & 3.51 & 0.55  &  1.59 & 0.09 \\
BD+27645    & 4:11:16.4 & 27:55:45.0 &  10.1 & 6.07 & 2.59  &  1.02 & 0.09 \\
PPM\,93320  & 4:11:57.1 & 27:10:05.9 &  10.3 & 3.31 & 1.12  &  0.66 & 0.10 \\
BD+24636    & 4:13:53.5 & 25:02:09.5 &  10.4 & 2.75 & 0.39  &  1.43 & 0.09 \\
PPM\,93369  & 4:15:24.6 & 29:21:57.1 &  10.2 & 4.11 & 0.59  &  2.05 & 0.14 \\
PPM\,93376  & 4:16:10.2 & 25:31:04.6 &   9.4 & 3.50 & 0.35  &  1.93 & 0.10 \\
PPM\,93377  & 4:16:11.1 & 29:07:15.4 &   9.7 & 2.93 & 0.27  &  1.98 & 0.09 \\
BD+25689    & 4:16:26.0 & 25:30:42.9 &  11.2 & 3.27 & 0.96  &  1.80 & 0.15 \\
PPM\,93390  & 4:17:13.3 & 27:19:44.5 &  11.1 & 6.70 & 2.34  &  2.01 & 0.12 \\
PPM\,93403  & 4:18:14.8 & 29:16:06.1 &  10.4 & 3.54 & 0.56  &  1.78 & 0.10 \\
BD+25698    & 4:18:46.2 & 26:08:57.1 &   9.2 & 2.74 & 0.29  &  1.36 & 0.08 \\
HD\,283581  & 4:20:07.4 & 26:24:40.5 &  11.4 & 2.31 & 1.26  &  0.70 & 0.16 \\
HD\,283569  & 4:20:48.3 & 28:29:39.6 &  11.2 & 2.32 & 0.52  &  1.52 & 0.13 \\
PPM\,93449  & 4:21:55.7 & 29:39:01.0 &   8.5 & 3.28 & 0.37  &  1.30 & 0.08 \\
PPM\,93510  & 4:25:33.0 & 28:26:58.1 &  10.5 & 1.62 & 0.40  &  0.77 & 0.08 \\
HD\,283625  & 4:26:51.6 & 28:57:11.1 &  11.4 & 2.95 & 0.92  &  1.71 & 0.15 \\
HD\,28170   & 4:27:34.0 & 25:03:41.7 &   9.0 & 3.30 & 0.32  &  1.62 & 0.08 \\
PPM\,93537  & 4:28:17.9 & 27:46:50.5 &   7.8 & 2.99 & 0.26  &  1.49 & 0.08 \\
PPM\,93546  & 4:29:02.9 & 26:30:58.9 &  10.8 & 2.38 & 0.38  &  1.65 & 0.11 \\
HD\,28482   & 4:30:22.4 & 23:35:19.9 &   7.2 & 3.87 & 0.50  &  1.67 & 0.14 \\
HD\,28975   & 4:34:50.2 & 24:14:40.3 &   9.0 & 3.22 & 0.28  &  1.72 & 0.07 \\
BD+26728    & 4:34:55.0 & 27:12:11.3 &   9.6 & 2.92 & 0.25  &  2.68 & 0.14 \\
PPM\,93637  & 4:37:09.1 & 27:55:32.7 &   7.5 & 3.55 & 0.76  &  0.68 & 0.08 \\
PPM\,93641  & 4:37:38.9 & 23:46:56.5 &  10.0 & 3.12 & 0.81  &  0.89 & 0.06 \\
PPM\,93642  & 4:37:46.3 & 24:02:45.9 &   9.7 & 3.19 & 0.38  &  1.56 & 0.09 \\
PPM\,93644  & 4:38:06.6 & 22:37:25.8 &   9.9 & 3.24 & 0.76  &  0.81 & 0.08 \\
HD\,29334   & 4:38:09.8 & 24:33:13.3 &   9.1 & 3.80 & 0.41  &  1.56 & 0.07 \\
BD+27675    & 4:38:15.2 & 27:52:50.7 &  10.7 & 3.34 & 0.81  &  1.18 & 0.09 \\
BD+22723    & 4:39:00.2 & 22:40:13.8 &  10.5 & 2.68 & 0.55  &  1.03 & 0.10 \\
PPM\,93658  & 4:39:06.7 & 22:42:43.4 &   9.8 & 2.90 & 0.47  &  0.97 & 0.09 \\
PPM\,93660  & 4:39:13.5 & 22:39:08.1 &   8.6 & 3.63 & 1.00  &  0.77 & 0.14 \\
PPM\,93668  & 4:39:59.1 & 23:00:52.5 &   8.8 & 4.33 & 0.47  &  1.67 & 0.08 \\
PPM\,93675  & 4:40:21.0 & 25:03:07.7 &   9.8 & 5.76 & 1.27  &  1.54 & 0.08 \\
HD\,283772  & 4:40:59.3 & 27:59:25.5 &  10.6 & 3.10 & 0.52  &  1.88 & 0.15 \\
BD+25724    & 4:42:19.9 & 25:51:48.3 &  10.8 & 3.76 & 0.56  &  2.61 & 0.12 \\
PPM\,93713  & 4:42:41.2 & 24:41:17.9 &  10.0 & 4.30 & 0.40  &  2.63 & 0.09 \\
BD+22741    & 4:42:44.0 & 22:36:19.4 &  10.9 & 6.29 & 2.41  &  1.55 & 0.12 \\
PPM\,93722  & 4:43:27.3 & 27:01:37.0 &   9.8 & 3.99 & 0.68  &  1.57 & 0.14 \\
BD+25727    & 4:44:24.9 & 25:31:42.7 &   9.5 & 3.20 & 0.28  &  2.12 & 0.09 \\
BD+26742    & 4:45:14.0 & 27:00:07.4 &  10.0 & 5.50 & 1.17  &  2.02 & 0.11 \\
HD\,30122   & 4:45:42.5 & 23:37:40.8 &   6.3 & 3.25 & 0.59  &  0.74 & 0.08 \\
BD+26746    & 4:46:02.8 & 26:18:39.6 &  10.0 & 3.50 & 0.50  &  1.71 & 0.09 \\
PPM\,93747  & 4:46:10.3 & 27:29:25.4 &  10.2 & 2.91 & 0.43  &  1.73 & 0.10 \\
HD\,30190   & 4:46:33.0 & 27:54:02.5 &   8.4 & 3.76 & 0.44  &  1.40 & 0.09 \\
HD\,283851  & 4:46:42.9 & 27:15:42.2 &  10.7 & 3.89 & 1.07  &  1.71 & 0.17 \\
PPM\,93771  & 4:47:27.7 & 24:21:17.5 &   9.9 & 3.98 & 0.68  &  1.42 & 0.09 \\
BD+27696    & 4:47:52.3 & 27:44:40.0 &   9.6 & 2.89 & 0.28  &  1.78 & 0.09 \\
PPM\,93776  & 4:47:54.1 & 26:33:38.4 &  10.3 & 2.85 & 0.54  &  1.25 & 0.10 \\
PPM\,93780  & 4:48:12.7 & 27:01:47.5 &  11.1 & 5.54 & 2.34  &  1.90 & 0.16 \\
PPM\,93819  & 4:50:58.5 & 24:16:42.8 &  11.0 & 4.38 & 1.57  &  1.66 & 0.14 \\
BD+25740    & 4:51:10.9 & 25:37:20.9 &  10.8 & 3.86 & 0.69  &  2.82 & 0.16 \\
PPM\,93854  & 4:53:09.5 & 25:29:28.0 &  10.3 & 5.45 & 1.38  &  1.74 & 0.10 \\
\enddata
\tablenotetext{a}{Apparent $V$-band magnitudes were obtained from the SIMBAD database,
operated at CDS, Strasbourg, France.  References include \citet{tycho2000} and \citet{adolfsson1954}.}

\end{deluxetable*}
%
\begin{deluxetable*}{llccccccccr}
\tabletypesize{\scriptsize}
\tablecaption{Stellar Target Sample for Optical Spectropolarimetry 
\label{sourcetab}}
\tablewidth{0pt}
\tablehead{
\colhead{} & \colhead{Alternate} & \colhead{RA (J2000)} & \colhead{Dec.\ (J2000)} & \colhead{$V$\tablenotemark{b}} & \colhead{Spectral} & \colhead{$A_V$\tablenotemark{d}} & \colhead{$R_V$\tablenotemark{e}} & \colhead{$R_V$\tablenotemark{f}} & \colhead{} & \colhead{Distance\tablenotemark{h}}\\
\colhead{Star\tablenotemark{a}} & \colhead{Name} & \colhead{(hh:mm:ss.s)} & \colhead{(dd:mm:ss.s)} & \colhead{(mag.)} & \colhead{Class\tablenotemark{c}} & \colhead{(mag.)} & \colhead{(mag.)} & \colhead{(mag.)} & \colhead{Source\tablenotemark{g}} & \colhead{(pc)} }%
\startdata
SWIW\,002 &  &  04:09:01.4 & +24:53:21.4 & 12.9 & K3  &  5.14$\pm$0.6 & 4.0$\pm$0.3 & 3.7$\pm$0.1 & A & 1030$_{-122}^{+158}$ \\
SWIW\,014 &  & 04:13:06.6 & +22:35:36.5 & 13.4 & M3  &  2.1$\pm$0.6 & 4.6$\pm$0.9 & 3.9$\pm$0.2 & A & 2658$_{-417}^{+568}$ \\ 
SWIW\,019  &  & 04:13:48.7 & +28:23:43.6 & 14.0 & K7  &  4.6$\pm$0.5 & 5.3$\pm$0.4 & 4.9$\pm$0.2 & A &2128$_{-294}^{+394}$ \\
SWIW\,026 &  & 04:15:24.1 & +28:07:07.4 & 14.7 & F8  &  5.6$\pm$0.5 & 5.0$\pm$1.4 & 4.7$\pm$0.2 & A &158$_{-1}^{+1}$ \\ 
SWIW\,040 & V409 Tau & 04:18:10.8 & +25:19:57.4 & 12.5 & M0e &  5.5$\pm$0.6 & \nodata & 8.0$\pm$3.1 & A & 131$_{-1}^{+1}$\\ 
SWIW\,046 &  & 04:19:58.3 & +28:12:13.9 & 14.3 & K4  &  4.5$\pm$0.5 & 5.3$\pm$0.1 & 4.9$\pm$0.1 & A & 1713$_{-217}^{+285}$ \\
SWIW\,049  & & 04:20:41.4 & +27:05:47.4 & 15.4 & G9  &  6.7$\pm$0.6 & 5.1$\pm$0.1 & 4.8$\pm$0.1 & S & 614$_{-28}^{+31}$ \\
SWIW\,051 &  & 04:21:00.0 & +30:22:17.9 & 13.8 & M5  &  1.3$\pm$0.8 & 8.6$\pm$0.2 & 7.4$\pm$0.6 & S(A) & 2314$_{-520}^{+808}$ \\
SWIW\,057 &  & 04:23:17.8 & +28:06:26.0 & 13.7 & M4  &  2.8$\pm$0.8 & 4.1$\pm$0.6 & 3.6$\pm$0.2 & A & 3355$_{-710}^{+1015}$ \\
SWIW\,093 &  & 04:30:38.7 & +22:55:52.0 & 17.1 & K4  &  9.1$\pm$0.8 & 6.4$\pm$0.1 & 6.2$\pm$0.2 & S & 574$_{-34}^{+38}$\\
SWIW\,100  & JH 57 & 04:31:26.4 & +27:07:20.4 & 14.7 & F0  &  6.0$\pm$0.5 & 2.3$\pm$0.2 & 2.1$\pm$0.1 & A & 164$_{-3}^{+3}$ \\
SWIW\,101 &  & 04:31:31.6 & +24:39:42.4 & 12.8 & K3  &  4.2$\pm$0.6 & 4.0$\pm$0.4 & 3.6$\pm$0.1 & A & 966$_{-82}^{+98}$ \\
SWIW\,109 &  & 04:32:01.3 & +28:13:34.7 & 13.5 & K7  &  3.0$\pm$0.6 & 3$\pm$2 & 2.5$\pm$0.1 & A & 1973$_{-296}^{+409}$ \\
SWIW\,121 & CoKu HK Tau G1  & 04:32:41.7 & +24:19:03.8 & 16.1 & F0  &  8.2$\pm$0.6 & 3.3$\pm$0.1 & 3.3$\pm$0.2 & A & 140$_{-3}^{+3}$ \\
SWIW\,125 & JH 114 & 04:33:21.6 & +22:39:50.4 & 13.4 & K1  &  4.4$\pm$0.5 & 12.2$\pm$0.6 & 11.0$\pm$0.4 & P & 476$_{-16}^{+17}$ \\
SWIW\,144 &  & 04:34:38.5 & +22:42:13.3 & 13.2 & K1  &  4.8$\pm$0.6 & 5.3$\pm$0.1 & 4.9$\pm$0.1 & P & 380$_{-11}^{+11}$ \\
SWIW\,148 & HO Tau & 04:35:20.2 & +22:32:14.6 & 14.5 & M3e & \nodata & \nodata &  \nodata &  & 161$_{-1}^{+1}$ \\
SWIW\,158 &  & 04:36:30.0 & +23:18:38.3 & 13.7 & M2  &  3.6 $\pm$0.9 & 2.7$\pm$0.1 & 2.4$\pm$0.1 & A & 2138$_{-378}^{+547}$\\
SWIW\,159 &  & 04:36:35.1 & +25:26:42.5 & 13.5 & G7  &  4.9$\pm$0.9 & 5.7$\pm$0.1 & 5.1$\pm$0.1 & A & 823$_{-63}^{+74}$ \\
SWIW\,163 &  & 04:37:13.7 & +24:22:20.8 & 13.5 & K7  &  3.2$\pm$0.5 & 3.9$\pm$1.0 &  3.5$\pm$0.1 & A & 1710$_{-242}^{+330}$\\
SWIW\,184 & JH 214 & 04:39:07.0 & +26:27:19.9 & 15.7 & F0  &  6.2$\pm$0.6 & 2.8$\pm$0.1 & 2.7$\pm$0.1 & S & 287$_{-9}^{+9}$\\
HD\,283809 & & 04:41:24.9 & +25:54:48.0 & 10.9 & B3  & 5.3 & 3.5$\pm$0.2 & 3.3$\pm$0.1 & A & 323$_{-9}^{+10}$ \\ 
SWIW217 & Kim 1-69, JH 227 & 04:42:35.7 & +25:27:15.2 & 12.8 & K4 & 5.6$\pm$0.8 & 4.5$\pm$0.9 & 4.3$\pm$0.1 & A & 512$_{-31}^{+35}$ \\
SWIW\,230 &  & 04:43:48.7 & +24:57:30.6 & 13.2 & K7  &  3.8$\pm$0.5 & 4.6$\pm$0.8 & 4.2$\pm$0.1 & A & 979$_{-121}^{+159}$ \\      
Tamura\,17 & Kim 1-88  & 04:44:01.5 & +25:20:13.0 & 11.1 & M5 & 2.4$\pm$0.8 & 7.1$\pm$1.8 & 7.2$\pm$1.5 & A & 379$_{-20}^{+23}$
\enddata
\tablenotetext{a}{SWIW refers to star numbers in the catalog of \citet{shenoy2008}.}
\tablenotetext{b}{$V$-band brightnesses from the NOMAD$^2$ compilation.}
\tablenotetext{c}{Spectral classes estimated from this work (see
  Section~\ref{sec-kast}). Uncertainties on all classes are 1--2
  subclasses.}
\tablenotetext{d}{Extinction values from \citet{shenoy2008}.}
\tablenotetext{e}{$R_V=1.1 \cdot E(V-K)/{E(B-V)}$ based on the spectral classes given in column 6 and photometry from AAVSO \citep{henden2016}, SDSS, and 2MASS}
\tablenotetext{f}{R$_V$ based on the spectral classes given in column 6 and fits of E($\lambda-V$)/E($B-V$) using data from AAVSO \citep{henden2016}, SDSS, 2MASS, and WISE}
\tablenotetext{g}{Source of visual photometry in order of preference: A: AAVSO, S: SDSS, P: Panstarrs.  For SWIW051, while AAVSO data exist, the resulting fits are poor.}
\tablenotetext{h}{From Gaia DR2}

\end{deluxetable*}

\section{Results} \label{sec-results}

\subsection{Polarization Fits} \label{sec-fits}

Stokes parameters $q$ and $u$, polarization amplitudes $p$, position
angles $\theta$, and uncertainties on each value, at each wavelength,
are listed in Table~\ref{tbl-results-turpol} for the low-extinction
sample and in Table~\ref{tbl-results} for the high-extinction star
sample.  The polarization amplitudes and angle uncertainties have been
corrected for positive noise bias
\citep[e.g.,][]{wardle1974,simmons1985,vaillancourt2006}.  For data
with signal to noise in the range $p/\sigma_p > \sqrt{2}$ we use
$p_\mathrm{corr} = (p^2-\sigma_p^2)^{1/2}$ and
$\sigma_\theta = 26.8\arcdeg\times(\sigma_p/p_\mathrm{corr})$.  For
$p/\sigma_p \le \sqrt{2}$ we set $p_\mathrm{corr} = 0$; in this case
angle measurements are not meaningful so neither angles nor their
uncertainties are listed.

\begin{figure}
  \plotone{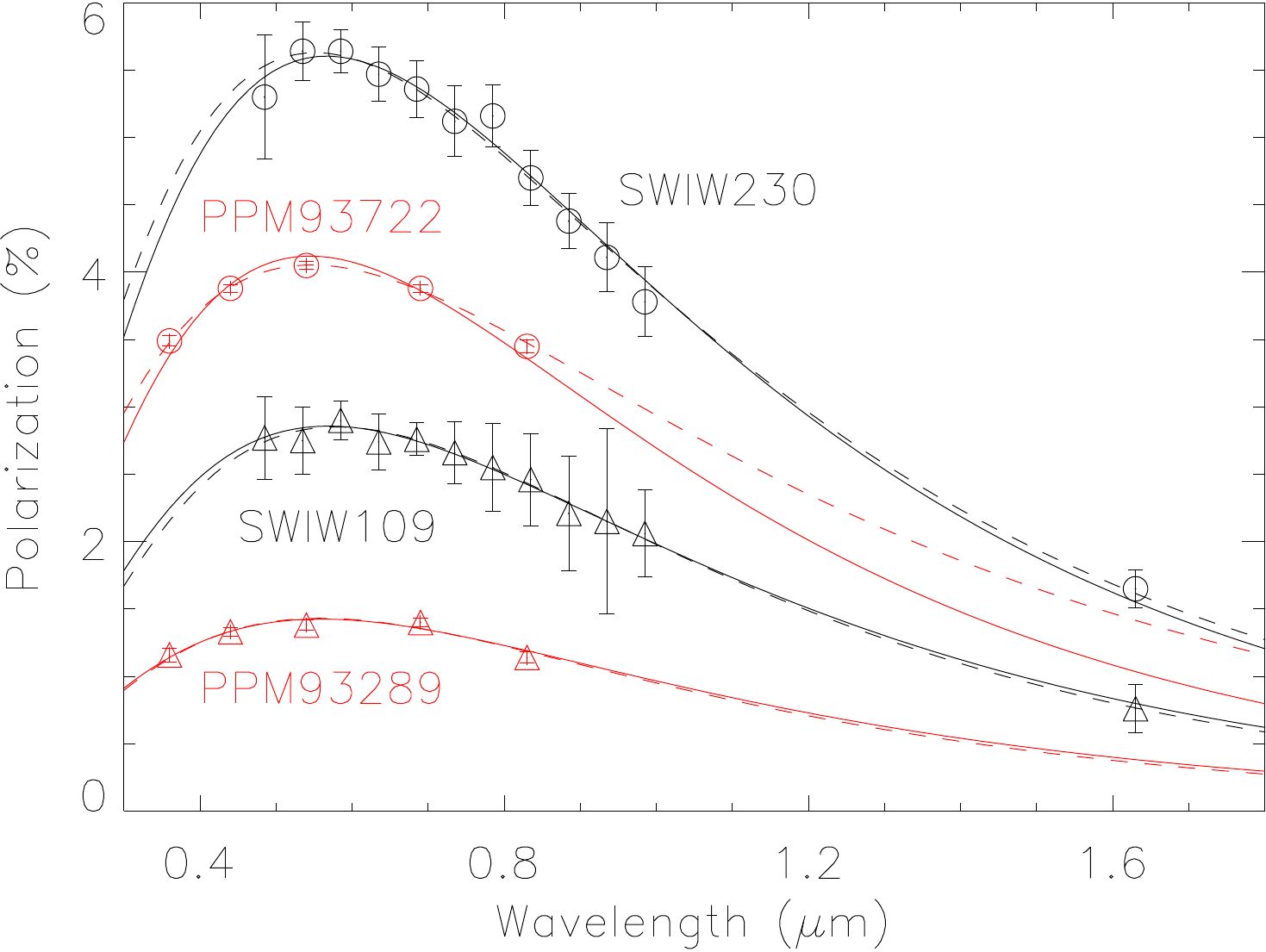}
  \caption{Example fits to the Serkowski curve for four stars. Black
    circles and triangles represent data for two high-extinction stars
    at 11 optical wavelengths bins and $H$-band, while the red points
    represent low-extinction stars at the $U\! BV\! RI$ passbands
    only. Solid lines represent fits to the Serkowski relation (with
    parameter $K=1.15$) whereas $K$ is allowed to float in the
    dashed-line fits to the Wilking relation.}
\label{fig-serk_ex1}
\end{figure}

As can be seen in Figure~\ref{fig-serk_ex1} (and gleaned from
Tables~\ref{tbl-serkowskifits-turpol} and \ref{tbl-serkowskifits}),
the polarization for most objects follows the expected form (equation
\ref{serkowski}), both for the case of fixing the parameter $K=1.15$
(the ``Serkowski'' relation; \citealt{serkowski1975}) and when $K$ is
a fitted quantity (the ``Wilking'' relation;
\citealt{wilking1980}). The change in the resulting goodness-of-fit
(as measured using a standard reduced-$\chi^2$;
Tables~\ref{tbl-serkowskifits-turpol} and \ref{tbl-serkowskifits})
between these different options for $K$ is marginal at best (the
quoted $\chi^2$ corresponds to a ``Wilking fit'' if a $K$-value is
quoted, otherwise to a ``Serkowski fit'').  To check whether the
additional term in the fit is statistically justified, we performed
$F$-tests by calculating the quantity $F_\chi$
\citep[e.g.,][]{bevington1992}. As reported in
Tables~\ref{tbl-serkowskifits-turpol} and \ref{tbl-serkowskifits},
only 11 low-extinction stars and three high-extinction stars yield
$F_\chi > 5$ (i.e., a $>$94\% probability that the additional term is
justified).

%
%
\begin{deluxetable*}{lccccccccc}
\tablewidth{0pt}
\tablecaption{Photo-polarimetry Results from the Low-extinction Sample \label{tbl-results-turpol}}
\tablehead{%
\colhead{} & \colhead{Wavelength\tablenotemark{a}} & 
\colhead{$q$} & \colhead{$\sigma_q$} & \colhead{$u$} & \colhead{$\sigma_u$} &
\colhead{$p_\mathrm{corr}$\tablenotemark{b}} & \colhead{$\sigma_p$} & 
\colhead{$\theta$\tablenotemark{c}} & \colhead{$\sigma_\theta$\tablenotemark{b,c}}
 \\
\colhead{Star} & \colhead{($\micron$)} & \colhead{(\%)} & \colhead{(\%)} &
\colhead{(\%)} & \colhead{(\%)} &\colhead{(\%)} & \colhead{(\%)} &
\colhead{(deg.)} & \colhead{(deg.)}
}

\startdata
PPM93195 &  0.36 &  $-$0.912 &  0.050 &   0.912 &  0.050 &  1.29 &  0.05 &  67.5  &  1.1 \\
PPM93195 &  0.44 &  $-$1.104 &  0.039 &   1.044 &  0.039 &  1.52 &  0.04 &  68.3  &  0.8 \\
PPM93195 &  0.55 &  $-$1.226 &  0.038 &   1.074 &  0.037 &  1.63 &  0.04 &  69.4  &  0.7 \\
PPM93195 &  0.69 &  $-$1.111 &  0.029 &   1.081 &  0.029 &  1.55 &  0.03 &  67.9  &  0.6 \\
PPM93195 &  0.83 &  $-$1.078 &  0.030 &   0.924 &  0.030 &  1.42 &  0.03 &  69.7  &  0.6 \\
\vdots & \vdots & \vdots & \vdots & \vdots & \vdots & \vdots & \vdots & \vdots & \vdots \\
PPM93236 &  0.36 &  $-$0.007 &  0.017 &   0.019 &  0.020 &  0.00 &  0.02 &    \nodata &     \nodata \\
PPM93236 &  0.44 &  $-$0.097 &  0.019 &   0.071 &  0.019 &  0.12 &  0.02 &   71.9 &  4.8\phn \\
PPM93236 &  0.55 &  $-$0.150 &  0.020 &   0.002 &  0.025 &  0.15 &  0.02 &   89.6 &  3.9\phn \\
PPM93236 &  0.69 &  $-$0.121 &  0.020 &   0.047 &  0.020 &  0.13 &  0.02 &   79.3 &  4.5\phn \\
PPM93236 &  0.83 &  $-$0.074 &  0.031 &   0.067 &  0.032 &  0.10 &  0.03 &   69.0 &  9.0\phn \\
\enddata

\tablenotetext{a}{Wavelength centers for broad-band filters $U$, $B$,
  $V$, $R$, and $I$ at NOT/Turpol are assigned as 0.36, 0.44, 0.55,
  0.69, and $0.83\,\micron$, respectively. Note these are
  not the same as the $V\! RI$-like bands defined for the Kast/Lick data (Appendix \ref{sec-pcal}; Table \ref{tbl-8}).}

\tablenotetext{b}{Polarization amplitude and angle uncertainty are
  corrected for positive noise bias (Section~\ref{sec-fits}).}

\tablenotetext{c}{All angles are measured east of north. No values are
  reported for $\theta$ and $\sigma_\theta$ for cases in which the
  corrected polarization is set to zero.}

\tablecomments{Polarization data for stars observed with the
  TurPol instrument on NOT\@. Listed uncertainties are statistical
  only and returned as part of the fitting procedure; uncertainties
  here do not include other systematics discussed in the text. (This
  table is available in its entirety in a machine-readable form in the
  online journal. A portion is shown here for guidance regarding its
  form and content.)}

\end{deluxetable*}
\begin{deluxetable*}{lcccccccccc}
\tablewidth{0pt}
\tablecaption{Spectropolarimetry and H-band Results for the High-extinction Sample\label{tbl-results}}
\tablehead{%
\colhead{} & \colhead{Wavelength\tablenotemark{a}} & 
\colhead{$q$} & \colhead{$\sigma_q$} & \colhead{$u$} & \colhead{$\sigma_u$} &
\colhead{$p_\mathrm{corr}$\tablenotemark{b}} & \colhead{$\sigma_p$} & 
\colhead{$\theta$\tablenotemark{c}} & \colhead{$\sigma_\theta$\tablenotemark{c}} &
\colhead{$\chi_r^2$\tablenotemark{d}}
 \\
\colhead{Star} & \colhead{($\micron$)} & \colhead{(\%)} & \colhead{(\%)} &
\colhead{(\%)} & \colhead{(\%)} &\colhead{(\%)} & \colhead{(\%)} &
\colhead{(deg.)} & \colhead{(deg.)} & \colhead{}
}

\startdata
 SWIW\,002 &  0.485 &   1.11 &  0.31 &  $-$3.77 &  0.31 &  3.92 &  0.31     &    143.2 &  2.2      &  0.79 \\
 SWIW\,002 &  0.535 &   0.80 &  0.19 &  $-$4.29 &  0.19 &  4.36 &  0.19     &    140.3 &  1.2      &  0.44 \\
\vdots &  \vdots &  \vdots &  \vdots &  \vdots &  \vdots &  \vdots & \vdots &  \vdots &  \vdots &  \vdots \\
 SWIW\,230 &  0.550 &   0.69 &  0.17 & \phs5.62 &  0.17 &  5.66 &  0.17     & \phn41.5 &  0.9      &  0.43 \\
 SWIW\,230 &  0.650 &   0.57 &  0.19 & \phs5.40 &  0.19 &  5.42 &  0.19     & \phn42.0 &  1.0      &  0.36 \\
 SWIW\,230 &  0.800 &   0.76 &  0.10 & \phs4.84 &  0.10 &  4.90 &  0.10     & \phn40.5 &  0.6      &  1.40 \\
 SWIW\,230 &  1.630 &   0.91 &  0.14 & \phs1.39 &  0.14 &  1.65 &  0.14  & \phn28.4 &  2.4   &  \nodata \\
\vdots &  \vdots &  \vdots &  \vdots &  \vdots &  \vdots &  \vdots & \vdots &  \vdots &  \vdots &  \vdots
\enddata

\tablenotetext{a}{Data at wavelengths 0.550, 0.650 and 0.800 $\micron$
  are binned centered at those wavelengths with full-widths of 0.100,
  0.100, and 0.150 $\micron$, respectively. Data at $1.630\,\micron$
  are $H$-band data. All other wavelength points are binned with
  widths of $0.050\,\micron$.}

\tablenotetext{b}{Polarization amplitude and angle uncertainty are
  corrected for positive noise bias (Section~\ref{sec-fits}).}

\tablenotetext{c}{All angles are measured east of north. No values are
  reported for $\theta$ and $\sigma_\theta$ for cases in which the
  corrected polarization is set to zero.}

\tablenotetext{d}{$\chi_r^2$ values are not reported for $H$-band data.}

\tablecomments{Polarization data for the high-extinction stars
  are listed here, with the exception of polarized standards (see
  Table~\ref{tbl-polstds}). Listed uncertainties are statistical only
  and returned as part of the fitting procedure along with the
  reduced-$\chi^2$ reported in the last column; uncertainties here do
  not include other systematics discussed in the text. (This table is
  available in its entirety in a machine-readable form in the online
  journal. A portion is shown here for guidance regarding its form and
  content.)}

\end{deluxetable*}
%
%
\clearpage
\startlongtable
\begin{deluxetable*}{lCCCCRCCDcDD}
\tablewidth{0pt}
\tablecaption{Fitted Polarization Parameters and Measured Angles for the Low-extinction Sample \label{tbl-serkowskifits-turpol}}
\tablehead{%
\colhead{} & \multicolumn{9}{c}{Serkowski ($K=1.15$) and Wilking fits} & \colhead{} & \multicolumn{4}{c}{Position Angles}\\
\cline{2-10}  \cline{12-15} \\
\colhead{} & 
\colhead{$p_\mathrm{max}$} & \colhead{$\sigma_{P_\mathrm{max}}$} &
\colhead{$\lambda_\mathrm{max}$} & \colhead{$\sigma_{\lambda_\mathrm{max}}$} &
\colhead{$K$} & \colhead{$\sigma_K$} &
\colhead{$\chi^2_r$} & \multicolumn{2}{c}{$F_\chi$\tablenotemark{c}} & 
\colhead{} &
\multicolumn2c{$\langle\theta\rangle$\tablenotemark{a}} & 
\multicolumn2c{$\sigma_{\langle\theta\rangle}$\tablenotemark{a}}
 \\
\colhead{Star} &
\colhead{(\%)} & \colhead{(\%)} &
\colhead{($\micron$)} & \colhead{($\micron$)} &
\colhead{} & \colhead{} & \colhead{} & 
\multicolumn2c{} &
\colhead{} & 
 \multicolumn2c{(deg.)} &
\multicolumn2c{(deg.)}
}
\decimals
\startdata
PPM\,93181	&	0.87	&	0.02	&	0.59	&	0.02	&	\nodata	&	\nodata	&	0.54	&	0.2	&&	98.8	&	1.5	\\
PPM\,93195	&	1.64	&	0.02	&	0.57	&	0.01	&	\nodata	&	\nodata	&	0.41	&	2.1	&&	68.7	&	1.1	\\
PPM\,93213	&	0.72	&	0.04	&	0.48	&	0.04	&	\nodata	&	\nodata	&	0.4	&	0.4	&&	127.7	&	4.6	\\
PPM\,93236	&	0.15	&	0.02	&	0.57	&	0.04	&	3.29	&	2.26	&	0.04	&	25.1	&&	78.6	&	12.1	\\
PPM\,92238	&	1.2	&	0.01	&	0.57	&	0.01	&	\nodata	&	\nodata	&	0.33	&	1.3	&&	104.1	&	1.5	\\
PPM\,93241	&	0.22	&	0.03	&	0.45	&	0.11	&	\nodata	&	\nodata	&	1.51	&	\nodata	&&	51.8	&	11.4	\\
PPM\,93260	&	0.67	&	0.01	&	0.62	&	0.02	&	\nodata	&	\nodata	&	3.23	&	0.2	&&	105.3	&	0.9	\\
PPM\,93265	&	0.95	&	0.03	&	0.72	&	0.03	&	\nodata	&	\nodata	&	1.01	&	0.1	&&	23.4	&	2.5	\\
PPM\,93280	&	3.58	&	0.03	&	0.62	&	0.01	&	\nodata	&	\nodata	&	0.59	&	6.6	&&	78.4	&	1	\\
PPM\,93281	&	0.59	&	0.05	&	0.54	&	0.02	&	2.89	&	0.94	&	0.92	&	3.7	&&	69.9	&	3.9	\\
PPM\,93289	&	1.43	&	0.02	&	0.56	&	0.01	&	\nodata	&	\nodata	&	1.76	&	0	&&	164.1	&	2.2	\\
BD+27645  	&	1.12	&	0.02	&	0.55	&	0.01	&	\nodata	&	\nodata	&	1.28	&	2.5	&&	81	&	1.5	\\
PPM\,93320	&	0.66	&	0.03	&	0.59	&	0.04	&	\nodata	&	\nodata	&	0.81	&	1.6	&&	63	&	2.3	\\
BD+24636  	&	0.77	&	0.04	&	0.63	&	0.05	&	\nodata	&	\nodata	&	0.44	&	0	&&	19.4	&	2.8	\\
PPM\,93369	&	3.4	&	0.02	&	0.53	&	0.01	&	\nodata	&	\nodata	&	7.07	&	3.6	&&	157.2	&	0.4	\\
PPM\,93376	&	2.83	&	0.02	&	0.68	&	0.01	&	\nodata	&	\nodata	&	2.02	&	0.1	&&	108.7	&	1.5	\\
PPM\,93377	&	3.11	&	0.02	&	0.60	&	0.01	&	0.83	&	0.08	&	0.1	&	163.9	&&	157.2	&	0.6	\\
BD+25689  	&	2.94	&	0.04	&	0.64	&	0.01	&	\nodata	&	\nodata	&	1.33	&	0.3	&&	104.1	&	0.7	\\
PPM\,93390	&	2.4	&	0.03	&	0.60	&	0.01	&	\nodata	&	\nodata	&	0.87	&	2.8	&&	57	&	0.7	\\
PPM\,93403	&	1.88	&	0.02	&	0.52	&	0.01	&	\nodata	&	\nodata	&	2.29	&	1.6	&&	172.1	&	1.5	\\
DB+25698  	&	0.3	&	0.01	&	0.51	&	0.03	&	\nodata	&	\nodata	&	3.04	&	42.7	&&	153	&	3.6	\\
HD\,283581	&	0.99	&	0.04	&	0.59	&	0.04	&	\nodata	&	\nodata	&	3.37	&	0.6	&&	41.6	&	2.7	\\
HD\,283569	&	2.66	&	0.05	&	0.53	&	0.03	&	0.46	&	0.18	&	0.43	&	32.7	&&	14	&	1	\\
PPM\,93449	&	0.61	&	0.02	&	0.55	&	0.03	&	\nodata	&	\nodata	&	0.65	&	0.5	&&	10.2	&	2	\\
PPM\,93510	&	2.25	&	0.04	&	0.50	&	0.02	&	0.64	&	0.16	&	0.37	&	26.5	&&	179.9	&	263.7	\\
HD\,283625	&	1.61	&	0.05	&	0.53	&	0.03	&	\nodata	&	\nodata	&	2.07	&	0.4	&&	177.9	&	6	\\
HD\,28170 	&	1.97	&	0.02	&	0.56	&	0.01	&	\nodata	&	\nodata	&	0.71	&	0.4	&&	88	&	0.7	\\
PPM\,93537	&	2	&	0.01	&	0.55	&	0.01	&	\nodata	&	\nodata	&	2.23	&	5.3	&&	12.4	&	0.6	\\
PPM\,93546	&	1.42	&	0.05	&	0.59	&	0.02	&	1.81	&	0.47	&	0.2	&	9.8	&&	26	&	1.7	\\
HD\,28482 	&	1.99	&	0.02	&	0.57	&	0.01	&	\nodata	&	\nodata	&	1.52	&	0.3	&&	55.7	&	2.9	\\
HD\,28975 	&	3.39	&	0.01	&	0.55	&	0.01	&	\nodata	&	\nodata	&	3.25	&	4.4	&&	58.5	&	0.8	\\
BD+26728  	&	3.31	&	0.02	&	0.58	&	0.01	&	\nodata	&	\nodata	&	1.2	&	0	&&	33.9	&	0.6	\\
PPM\,93637	&	1.69	&	0.01	&	0.54	&	0.01	&	\nodata	&	\nodata	&	2.48	&	1.7	&&	33.1	&	0.3	\\
PPM\,93641	&	1.2	&	0.03	&	0.55	&	0.01	&	1.72	&	0.29	&	0.3	&	12.9	&&	71.5	&	1.9	\\
PPM\,93642	&	1.71	&	0.02	&	0.57	&	0.01	&	\nodata	&	\nodata	&	0.23	&	1.1	&&	69.4	&	2.2	\\
PPM\,93644	&	2.65	&	0.03	&	0.53	&	0.01	&	0.92	&	0.1	&	0.14	&	35.8	&&	60.3	&	0.4	\\
HD\,29334 	&	1.83	&	0.02	&	0.52	&	0.01	&	\nodata	&	\nodata	&	2.17	&	0.1	&&	44.9	&	0.9	\\
BD+27675  	&	1.81	&	0.02	&	0.54	&	0.01	&	\nodata	&	\nodata	&	1.6	&	0.8	&&	29.7	&	0.9	\\
BD+22723  	&	2.24	&	0.03	&	0.50	&	0.01	&	\nodata	&	\nodata	&	1.22	&	6.8	&&	60.3	&	0.9	\\
PPM\,93658	&	2.41	&	0.02	&	0.53	&	0.01	&	\nodata	&	\nodata	&	1.11	&	2.9	&&	56	&	0.6	\\
PPM\,93660	&	2.24	&	0.01	&	0.52	&	0.01	&	\nodata	&	\nodata	&	4.3	&	3	&&	54.6	&	1	\\
PPM\,93668	&	1.03	&	0.01	&	0.65	&	0.01	&	\nodata	&	\nodata	&	4	&	0	&&	41.5	&	0.9	\\
PPM\,93675	&	0.7	&	0.02	&	0.60	&	0.03	&	\nodata	&	\nodata	&	0.35	&	0.1	&&	42.8	&	2.9	\\
HD\,283772	&	0.61	&	0.03	&	0.62	&	0.04	&	\nodata	&	\nodata	&	2.09	&	7	&&	96.6	&	7.2	\\
BD+25724  	&	5.43	&	0.03	&	0.55	&	0.01	&	\nodata	&	\nodata	&	1.13	&	3.2	&&	39.2	&	1.2	\\
PPM\,93713	&	2.88	&	0.02	&	0.55	&	0.01	&	0.95	&	0.07	&	1.42	&	5.1	&&	43.8	&	0.3	\\
BD+22741  	&	2.32	&	0.03	&	0.51	&	0.01	&	\nodata	&	\nodata	&	0.57	&	0.3	&&	61.4	&	0.8	\\
PPM\,93722	&	4.05	&	0.02	&	0.55	&	0.01	&	0.88	&	0.07	&	0.38	&	43.6	&&	24.9	&	1	\\
BD+25727  	&	6.36	&	0.01	&	0.57	&	0.01	&	\nodata	&	\nodata	&	8.13	&	0.1	&&	31.8	&	0.3	\\
BD+26742  	&	3.75	&	0.02	&	0.55	&	0.01	&	\nodata	&	\nodata	&	0.84	&	1.9	&&	32.5	&	1.1	\\
HD\,30122 	&	1.26	&	0.02	&	0.53	&	0.01	&	\nodata	&	\nodata	&	0.85	&	3	&&	61.2	&	0.6	\\
BD+26746  	&	4.67	&	0.02	&	0.55	&	0.01	&	\nodata	&	\nodata	&	2.93	&	0.2	&&	26.8	&	0.3	\\
PPM\,93747	&	3.48	&	0.02	&	0.54	&	0.01	&	\nodata	&	\nodata	&	3.44	&	0.7	&&	40.8	&	0.6	\\
HD\,30190 	&	3.7	&	0.02	&	0.55	&	0.01	&	\nodata	&	\nodata	&	1.18	&	0.4	&&	56.4	&	0.3	\\
HD\,283851	&	3.15	&	0.03	&	0.57	&	0.01	&	\nodata	&	\nodata	&	0.31	&	0.3	&&	40.9	&	0.6	\\
PPM\,93771	&	1.23	&	0.03	&	0.52	&	0.03	&	\nodata	&	\nodata	&	0.51	&	0.4	&&	51	&	2.1	\\
BD+27696  	&	4.04	&	0.02	&	0.53	&	0.01	&	\nodata	&	\nodata	&	12.29	&	3.4	&&	59.5	&	0.9	\\
PPM\,93776	&	3.62	&	0.02	&	0.55	&	0.01	&	\nodata	&	\nodata	&	1.39	&	0	&&	33.3	&	0.7	\\
PPM\,93780	&	5.17	&	0.03	&	0.53	&	0.01	&	\nodata	&	\nodata	&	1.64	&	8.7	&&	44.9	&	0.4	\\
PPM\,93819	&	1.29	&	0.04	&	0.46	&	0.02	&	\nodata	&	\nodata	&	3.1	&	0.6	&&	49.4	&	1.6	\\
BD+25740  	&	3.52	&	0.03	&	0.55	&	0.01	&	\nodata	&	\nodata	&	1.84	&	0	&&	40.7	&	0.5	\\
PPM\,93854	&	2.67	&	0.03	&	0.61	&	0.01	&	\nodata	&	\nodata	&	0.44	&	1.3	&&	50.3	&	0.8	\\
 PPM\,93181	& 0.85	&  0.02	& 0.57 &  0.02	& \nodata & \nodata &    0.57  &    0.3 & &   98.8 &      1.5  \\
 PPM\,93195	& 1.62	&  0.02	& 0.56 &  0.01	& \nodata & \nodata &    0.48  &    0.5 & &   68.7 &      1.1  \\
 PPM\,93213	& 0.72	&  0.04	& 0.47 &  0.04	& \nodata & \nodata &    0.57  &    0.4 & &  127.7 &      4.6  \\
 PPM\,93236	& 0.14	&  0.02	& 0.56 &  0.04	&    3.49 &    2.41 &    0.11  &    8.3 & &   78.6 &     12.1  \\
 PPM\,92238	& 1.18	&  0.01	& 0.56 &  0.01	& \nodata & \nodata &    0.09  &    0.2 & &  104.1 &      1.5  \\
 PPM\,93241	& 0.22	&  0.03	& 0.43 &  0.12	& \nodata & \nodata &    1.67  &     \nodata  & &   51.8 &     11.4  \\
 PPM\,93260	& 0.67	&  0.01	& 0.61 &  0.02	& \nodata & \nodata &    3.16  &    0.1 & &  105.3 &      0.9  \\
 PPM\,93265	& 0.95	&  0.03	& 0.72 &  0.04	& \nodata & \nodata &    0.73  &    0.3 & &   23.4 &      2.5  \\
 PPM\,93280	& 3.54	&  0.03	& 0.61 &  0.01	& \nodata & \nodata &    0.43  &    2.5 & &   78.4 &      1.0  \\
 PPM\,93281	& 0.59	&  0.04	& 0.52 &  0.02	&    3.23 &    1.03 &    0.78  &    5.2 & &   69.9 &      3.9  \\
 PPM\,93289	& 1.41	&  0.02	& 0.54 &  0.01	& \nodata & \nodata &    1.31  &    0.3 & &  164.1 &      2.2  \\
  BD+27645	& 1.11	&  0.02	& 0.54 &  0.01	& \nodata & \nodata &    1.13  &    1.8 & &   81.0 &      1.5  \\
 PPM\,93320	& 0.65	&  0.03	& 0.58 &  0.04	& \nodata & \nodata &    0.86  &    3.2 & &   63.0 &      2.3  \\
  BD+24636	& 0.77	&  0.04	& 0.63 &  0.05	& \nodata & \nodata &    0.42  &    0.0 & &   19.4 &      2.8  \\
 PPM\,93369	& 3.37	&  0.02	& 0.51 &  0.01	& \nodata & \nodata &    3.07  &    3.1 & &  157.2 &      0.4  \\
 PPM\,93376	& 2.82	&  0.02	& 0.67 &  0.01	& \nodata & \nodata &    4.19  &    0.4 & &  108.7 &      1.5  \\
 PPM\,93377	& 3.09	&  0.02	& 0.59 &  0.01	&    0.87 &    0.08 &    0.18  &   62.9 & &  157.2 &      0.6  \\
  BD+25689	& 2.92	&  0.04	& 0.64 &  0.02	& \nodata & \nodata &    1.03  &    0.5 & &  104.1 &      0.7  \\
 PPM\,93390	& 2.37	&  0.03	& 0.59 &  0.01	& \nodata & \nodata &    0.71  &    1.8 & &   57.0 &      0.7  \\
 PPM\,93403	& 1.86	&  0.02	& 0.51 &  0.01	& \nodata & \nodata &    1.47  &    1.5 & &  172.1 &      1.5  \\
  BD+25698	& 0.30	&  0.01	& 0.50 &  0.03	& \nodata & \nodata &    2.73  &   32.9 & &  153.0 &      3.6  \\
 HD\,283581	& 0.98	&  0.04	& 0.59 &  0.04	& \nodata & \nodata &    3.54  &    0.4 & &   41.6 &      2.7  \\
 HD\,283569	& 2.65	&  0.04	& 0.52 &  0.02	&    0.49 &    0.20 &    0.42  &   26.8 & &   14.0 &      1.0  \\
 PPM\,93449	& 0.61	&  0.02	& 0.54 &  0.03	& \nodata & \nodata &    0.84  &    0.4 & &   10.2 &      2.0  \\
 PPM\,93510	& 2.24	&  0.03	& 0.49 &  0.02	&    0.67 &    0.17 &    0.46  &   17.6 & &  179.9 &      0.3  \\
 HD\,283625	& 1.60	&  0.05	& 0.52 &  0.03	& \nodata & \nodata &    2.41  &    0.4 & &  177.9 &      6.0  \\
  HD\,28170	& 1.94	&  0.02	& 0.55 &  0.01	& \nodata & \nodata &    1.06  &    1.0 & &   88.0 &      0.7  \\
 PPM\,93537	& 1.98	&  0.01	& 0.53 &  0.01	& \nodata & \nodata &    1.69  &    2.4 & &   12.4 &      0.6  \\
 PPM\,93546	& 1.40	&  0.05	& 0.58 &  0.02	&    1.92 &    0.50 &    0.13  &   17.9 & &   26.0 &      1.7  \\
  HD\,28482	& 1.97	&  0.02	& 0.56 &  0.01	& \nodata & \nodata &    1.78  &    0.0 & &   55.7 &      2.9  \\
  HD\,28975	& 3.35	&  0.01	& 0.54 &  0.01	& \nodata & \nodata &    1.46  &    3.3 & &   58.5 &      0.8  \\
  BD+26728	& 3.27	&  0.02	& 0.57 &  0.01	& \nodata & \nodata &    1.49  &    0.1 & &   33.9 &      0.6  \\
 PPM\,93637	& 1.67	&  0.01	& 0.53 &  0.01	& \nodata & \nodata &    0.60  &    1.2 & &   33.1 &      0.3  \\
PPM\,93641	& 1.19	&  0.03	& 0.54 &  0.01	&    1.81 &    0.31 &    0.84  &    5.4 & &   71.5 &      1.9  \\
 PPM\,93642	& 1.69	&  0.02	& 0.55 &  0.01	& \nodata & \nodata &    0.39  &    0.0 & &   69.4 &      2.2  \\
 PPM\,93644	& 2.63	&  0.02	& 0.52 &  0.01	&    0.98 &    0.11 &    0.15  &   15.4 & &   60.3 &      0.4  \\
  HD\,29334	& 1.81	&  0.02	& 0.50 &  0.01	& \nodata & \nodata &    0.54  &    0.1 & &   44.9 &      0.9  \\
  BD+27675	& 1.79	&  0.02	& 0.53 &  0.01	& \nodata & \nodata &    1.27  &    0.4 & &   29.7 &      0.9  \\
  BD+22723	& 2.22	&  0.03	& 0.49 &  0.01	& \nodata & \nodata &    0.79  &    2.3 & &   60.3 &      0.9  \\
 PPM\,93658	& 2.38	&  0.02	& 0.51 &  0.01	& \nodata & \nodata &    0.09  &    2.6 & &   56.0 &      0.6  \\
 PPM\,93660	& 2.23	&  0.01	& 0.51 &  0.01	& \nodata & \nodata &    2.35  &    0.5 & &   54.6 &      1.0  \\
 PPM\,93668	& 1.02	&  0.01	& 0.64 &  0.02	& \nodata & \nodata &    4.46  &    0.1 & &   41.5 &      0.9  \\
 PPM\,93675	& 0.69	&  0.02	& 0.59 &  0.03	& \nodata & \nodata &    0.35  &    0.4 & &   42.8 &      2.9  \\
 HD\,283772	& 0.61	&  0.03	& 0.61 &  0.04	& \nodata & \nodata &    1.98  &    7.3 & &   96.6 &      7.2  \\
  BD+25724	& 5.36	&  0.03	& 0.53 &  0.01	& \nodata & \nodata &    1.20  &    0.6 & &   39.2 &      1.2  \\
 PPM\,93713	& 2.86	&  0.02	& 0.54 &  0.01	&    0.98 &    0.08 &    0.28  &   17.5 & &   43.8 &      0.3  \\
  BD+22741	& 2.30	&  0.03	& 0.50 &  0.01	& \nodata & \nodata &    0.87  &    0.0 & &   61.4 &      0.8  \\
 PPM\,93722	& 4.04	&  0.02	& 0.53 &  0.01	&    0.96 &    0.07 &    0.36  &   19.1 & &   24.9 &      1.0  \\
  BD+25727	& 6.28	&  0.01	& 0.56 &  0.01	& \nodata & \nodata &    3.56  &    0.0 & &   31.8 &      0.3  \\
  BD+26742	& 3.70	&  0.02	& 0.54 &  0.01	& \nodata & \nodata &    0.48  &    0.7 & &   32.5 &      1.1  \\
  HD\,30122	& 1.27	&  0.02	& 0.51 &  0.01	&    1.59 &    0.24 &    0.23  &   14.8 & &   61.2 &      0.6  \\
  BD+26746	& 4.61	&  0.02	& 0.53 &  0.01	& \nodata & \nodata &    2.48  &    2.1 & &   26.8 &      0.3  \\
 PPM\,93747	& 3.44	&  0.02	& 0.53 &  0.01	& \nodata & \nodata &    1.39  &    0.3 & &   40.8 &      0.6  \\
  HD\,30190	& 3.66	&  0.02	& 0.54 &  0.01	& \nodata & \nodata &    0.14  &    0.0 & &   56.4 &      0.3  \\
 HD\,283851	& 3.11	&  0.03	& 0.56 &  0.01	& \nodata & \nodata &    0.30  &    0.1 & &   40.9 &      0.6  \\
 PPM\,93771	& 1.22	&  0.03	& 0.51 &  0.03	& \nodata & \nodata &    0.71  &    0.4 & &   51.0 &      2.1  \\
  BD+27696	& 4.00	&  0.02	& 0.51 &  0.01	& \nodata & \nodata &    5.64  &    2.0 & &   59.5 &      0.9  \\
 PPM\,93776	& 3.58	&  0.02	& 0.54 &  0.01	& \nodata & \nodata &    1.60  &    0.2 & &   33.3 &      0.7  \\
 PPM\,93780	& 5.11	&  0.03	& 0.52 &  0.01	& \nodata & \nodata &    0.57  &    1.7 & &   44.9 &      0.4  \\
 PPM\,93819	& 1.29	&  0.04	& 0.44 &  0.02	& \nodata & \nodata &    2.57  &    1.8 & &   49.4 &      1.6  \\
  BD+25740	& 3.47	&  0.03	& 0.54 &  0.01	& \nodata & \nodata &    1.95  &    0.0 & &   40.7 &      0.5  \\
 PPM\,93854	& 2.64	&  0.03	& 0.60 &  0.01	& \nodata & \nodata &    0.35  &    0.5 & &   50.3 &      0.8  \\
\enddata

\tablecomments{Parameters $p$, $\lambda_\mathrm{max}$, $K$ (Wilking),
  and their uncertainties are fitted to all stars, as are parameters
  where $K=1.15$ (Serkowski).  Here we report the three Wilking fit
  parameters only if an $F$-test of the extra term in the Wilking
  relation returns $F_\chi > 5$, otherwise the Serkowski values are
  reported. These fits are discussed in Appendix~\ref{sec-serkfits}.}

\tablenotetext{a}{$F_\chi \equiv (\chi^2_s - \chi^2_w) /
  (\chi^2_w/N)$, where $N$ is the number of degrees of freedom in the
  fits and $\chi_{s,w}$ are the $\chi^2_r$ reported in the table for
  the fits to the Serkowski and Wilking relations,
  respectively.}

\tablenotetext{b}{$\langle\theta\rangle$ is the variance-weighted mean
  (calculated in Stokes-space) of the measured position angles in the
  five broad-bands $U\! BV\! RI$ in this work. $\sigma_{\langle\theta\rangle}$
  is the larger of the median uncertainty or unweighted standard
  deviation of the five angles.}


\end{deluxetable*}
%
%
%
\startlongtable
\begin{deluxetable*}{lccccccccccccc}
\tablewidth{0pt}
\tablecaption{Fitted Polarization Parameters and Measured Angles for
  High-extinction Stars
  \label{tbl-serkowskifits}}
\tablehead{%
\colhead{} & \multicolumn{8}{c}{Serkowski ($K=1.15$) and Wilking fits} & \colhead{} & \multicolumn{4}{c}{Position Angles}\\
\cline{2-9}  \cline{11-14} \\
\colhead{} & 
\colhead{$p_\mathrm{max}$} & \colhead{$\sigma_{P_\mathrm{max}}$} &
\colhead{$\lambda_\mathrm{max}$} & \colhead{$\sigma_{\lambda_\mathrm{max}}$} &
\colhead{$K$} & \colhead{$\sigma_K$} &
\colhead{$\chi^2_r$} & \colhead{$F_\chi$\tablenotemark{a}} & 
\colhead{} &
\colhead{$\langle\theta\rangle$\tablenotemark{b}} &\colhead{$\sigma_{\langle\theta\rangle}$\tablenotemark{b}} & \colhead{$\theta(\mathrm{IR})$\tablenotemark{c}} & \colhead{$\sigma_{\theta(\mathrm{IR})}$}
 \\
\colhead{Star} &
\colhead{(\%)} & \colhead{(\%)} &
\colhead{($\micron$)} & \colhead{($\micron$)} &
\colhead{} & \colhead{} & \colhead{} & \colhead{} &
\colhead{} & 
\colhead{(deg.)} &\colhead{(deg.)} & \colhead{(deg.)} & \colhead{(deg.)}
}

\startdata
 &  \multicolumn{11}{c}{Fits include $H$-band data} \\
 SWIW\,002	&  5.29 & 0.08  & 0.762 & 0.015  &    1.53 &    0.22 &  0.05 &    61.7 & &    142.4	& \phn1.1 & 133.6       & \phn5.4  \\
 SWIW\,014	&  1.48	& 0.07 	& 0.679	& 0.065	 & \nodata & \nodata &  0.34 & \phn0.3 & &    111.5	& \phn4.4 & \phn89.7	&    21.2  \\
 SWIW\,019	&  2.81	& 0.07 	& 0.655	& 0.031	 & \nodata & \nodata &  0.20 & \phn1.8 & &    172.3	& \phn2.2 &    152.5	& \phn4.8  \\
 SWIW\,026	&  3.60	& 0.09 	& 0.896	& 0.035	 & \nodata & \nodata &  0.49 & \phn0.1 & & \phn21.1	& \phn4.9 & \phn\phn8.5	& \phn2.9  \\
 SWIW\,046	&  4.68	& 0.07 	& 0.610	& 0.015	 & \nodata & \nodata &  0.26 & \phn1.9 & & \phn\phn7.0	& \phn1.1 & \phn\phn4.5	& \phn2.3 \\
 SWIW\,049	&  2.56	& 0.08 	& 0.652	& 0.034	 & \nodata & \nodata &  0.95 & \phn2.6 & & \phn40.8	& \phn2.3\tablenotemark{d}
                                                                                                                  &  \phn32.1	& \phn4.3  \\
 SWIW\,057	&  6.60	& 0.13 	& 0.587	& 0.014	 & \nodata & \nodata &  2.05 & \phn0.6 & & \phn\phn1.7	& \phn1.3 &     171.3	& \phn2.1  \\
 SWIW\,093	&  1.44	& 0.14 	& 0.808	& 0.141	 & \nodata & \nodata &  0.50 & \phn0.2 & & \phn74.2	& \phn12.1&  \phn49.2	& \phn9.9  \\
 SWIW\,100	&  1.78	& 0.14 	& 1.131	& 0.092	 & \nodata & \nodata &  0.70 & \phn1.3 & & \phn20.2	& \phn5.3 &  \phn12.0	& \phn4.9  \\
 SWIW\,101	&  3.82	& 0.05 	& 0.712	& 0.020	 & \nodata & \nodata &  0.26 & \phn3.0 & & \phn76.7	& \phn1.5 &  \phn72.1	& \phn4.5  \\
 SWIW\,109	&  2.85	& 0.09 	& 0.567	& 0.035	 & \nodata & \nodata & 0.06 &  \phn1.1 & & \phn40.4	& \phn3.2 & \phn17.5	& \phn7.0  \\
 SWIW\,121\tablenotemark{e}
                &  3.08	& 0.71 	& 0.384	& 0.074	 & \nodata & \nodata & 0.31 &  \phn0.1 & & \phn85.3	&  19.9   &    139.2	&    14.4  \\
 SWIW\,125	&  2.41 & 0.07  & 0.585  & 0.068 & 0.81    & 0.30    & 0.20 &  \phn6.6 & & \phn53.8	& \phn2.1 & \phn40.8	& \phn6.1  \\
 SWIW\,144	&  2.03	& 0.06 	& 0.751	& 0.047	 & \nodata & \nodata & 0.37 &  \phn1.5 & & \phn26.5	& \phn5.7 & \phn\phn4.2	&    11.5  \\
 SWIW\,148	&  1.26	& 0.06 	& 0.628	& 0.061	 & \nodata & \nodata & 0.20 &  \phn0.0 & & \phn55.9	& \phn4.5 & \phn\phn7.4	&    18.4  \\
 SWIW\,158	&  5.14	& 0.09 	& 0.593	& 0.017	 & \nodata & \nodata & 0.12 &  \phn0.0 & & \phn51.5	& \phn1.6 & \phn35.8	& \phn6.9  \\
 SWIW\,159\tablenotemark{e}
                &  4.19 &  0.31 & 0.412 & 0.093  &  0.52   &    0.16 & 0.33 &     46.0 & & \phn44.0	& \phn2.3 & \phn27.1	& \phn2.7  \\
 SWIW\,163	&  4.24	& 0.08 	& 0.588	& 0.019	 & \nodata & \nodata & 0.58 &  \phn2.9 & & \phn52.5	& \phn1.5 & \phn52.5	& \phn4.9  \\
 SWIW\,184	&  2.11	& 0.12 	& 0.628	& 0.060	 & \nodata & \nodata & 0.30 &  \phn0.1 & & \phn55.7	& \phn4.1 & \phn48.1	& \phn8.1  \\
 SWIW\,230	&  5.60	& 0.10 	& 0.567	& 0.014	 & \nodata & \nodata & 0.24 &  \phn1.7 & & \phn41.2	& \phn1.4 & \phn28.4	& \phn2.4  \\
 &  \multicolumn{11}{c}{No $H$-band data} \\                                                 
 SWIW\,040\tablenotemark{e}
                &  6.50	& 0.28 	& 1.242	& 0.049	 & \nodata & \nodata & 0.38 &   \phn0.4	& & \phn50.1	& \phn2.4 & \phn49.7	& \phn0.4 \\
HD\,283809	&  6.68	& 0.14 	& 0.610	& 0.022	 & \nodata & \nodata & 0.01 &   \phn0.1	& & \phn53.6	& \phn2.5 & \phn54.5	& \phn0.4 \\
   Kim\,69	&  7.80	& 0.09 	& 0.623	& 0.014	 & \nodata & \nodata & 0.03 &   \phn0.4	& & \phn44.9	& \phn1.6 & \phn44.5	& \nodata \\
Tamura\,17	&  5.78	& 0.13 	& 0.555	& 0.017	 & \nodata & \nodata & 0.03 &   \phn0.7 & & \phn45.4	& \phn1.6 & \phn44.6	& \nodata \\
\enddata
\tablecomments{Parameters $p$, $\lambda_\mathrm{max}$, $K$ (Wilking),
  and their uncertainties are fitted to all stars, as are parameters
  where $K=1.15$ (Serkowski).  Here we report the three Wilking fit
  parameters only if an $F$-test of the extra term in the Wilking
  relation returns $F_\chi > 5$, otherwise the Serkowski values are
  reported. These fits are discussed in Appendix~\ref{sec-serkfits}.}

\tablenotetext{a}{$F_\chi \equiv (\chi^2_s - \chi^2_w) /
  (\chi^2_w/N)$, where $N$ is the number of degrees of freedom in the
  fits and $\chi_{s,w}$ are the $\chi^2_r$ reported in the table for
  the fits to the Serkowski and Wilking relations,
  respectively.}

\tablenotetext{b}{$\langle\theta\rangle$ is the variance-weighted mean
  (calculated in Stokes-space) of the measured position angles in the
  11 optical wavelengths in this work. $\sigma_{\langle\theta\rangle}$
  is the larger of the median uncertainty or unweighted standard
  deviation of the 11 angles.}

\tablenotetext{c}{$\theta(\mathrm{IR})$ is the position angle measured in
  $H$-band for all SWIW-objects.  For HD\,283809 this value is the
  mean angle at $\lambda = 0.35$--2.2 $\micron$ from
  \citet{whittet2001}.  For Kim\,69 and Tamura\,17 the angles are
  from the $K$-band measurements of \citet{tamura1987}.}

\tablenotetext{d}{Calculation of $\sigma_{\langle\theta\rangle}$ does
  not include outliers at 0.485 and 0.985~\micron.}

\tablenotetext{e}{The collected data for SWIW 040, 121, 159, and
  PPM\,93780 do not bracket a wavelength peak.  Therefore, the
  $p_\mathrm{max}$ and $\lambda_\mathrm{max}$ values here are likely
  unreliable.}


\end{deluxetable*}
%

Of the stars studied here, spectropolarimetry has previously been
performed for only HD\,283809, for which \citet{whittet2001} found
$p_\mathrm{max} = 6.70 \pm 0.10\%$,
$\lambda_\mathrm{max}= 0.59 \pm 0.02$ $\micron$, and
$K = 0.97 \pm 0.10$. Within the uncertainties, these values are
consistent with the values found here.

The stars SWIW\,121 and SWIW\,159 show no clear peak of polarization
with wavelength longward of the shortest measured wavelength of
$0.460\,\micron$. As a result, fits to find a peak ($p_\mathrm{max}$,
$\lambda_\mathrm{max}$) are unreliable. In SWIW\,121, this was most
likely due to very low signal-to-noise in the polarization signal and
is also reflected in the large $\lambda_\mathrm{max}$ uncertainty when
$K$ was allowed to float. The signal-to-noise was adequate for
SWIW\,159, so it is likely that $\lambda_\mathrm{max}$ is
intrinsically small.  A weighted Serkowski fit to the optical data
yields ($p_\mathrm{max}=3.87\%\pm0.08\%$,
$\lambda_\mathrm{max}=0.59\pm0.02\,\micron$, $\chi^2=3.1$). Data and
fit values for these two stars are not included in
Figure~\ref{fig-data_models}.

The Lick/Kast polarization data for SWIW\,040 continues to increase up
to the longest measured wavelength of $1.01\,\micron$ and neither the
Serkowski nor the Wilking fits yield a peak at shorter wavelengths.  A
weighted Serkowski fit to the optical data yields
($p_\mathrm{max}=6.45\%\pm0.30\%$,
$\lambda_\mathrm{max}=1.24\pm0.05\,\micron$, $\chi^2=3.6$).  Due to
the high S/N of this fit and the data, we include this star in the
analysis, and note that its inclusion alters the fit parameters of the
``upper branch'' by less than $2\sigma$ (Section~\ref{sec-discussion1}
and equation~[\ref{eq-linfit}]).

For SWIW\,051, all data, with the exception of the $0.685\,\micron$
bin and $H$-band, are consistent with zero polarization.  Therefore,
no fits are reported for that star in Table~\ref{tbl-serkowskifits}.

The spectra of the stars observed with Lick/Kast exhibit several
strong telluric absorption lines.  The molecular oxygen A-band
(0.760--0.763\,$\micron$) is by far the strongest of these and may
lead to contamination of the $0.785\,\micron$ bin.  This was
particularly evident for SWIW\,049 and for SWIW\,144, where the
polarization curve showed an unexpected ``dip'' compared to its
nearest spectral bins. However, removal of the A-band lines from the
$0.785\,\micron$ bin did not change this behavior and the change in
the resulting polarization (both amplitude and angle) was within the
uncertainties given in Table~\ref{tbl-results}. Therefore, we do not
expect any telluric absorption lines have a significant effect on the
polarization data reported here.

\subsection{Polarization Angles}

\begin{figure}
  \plotone{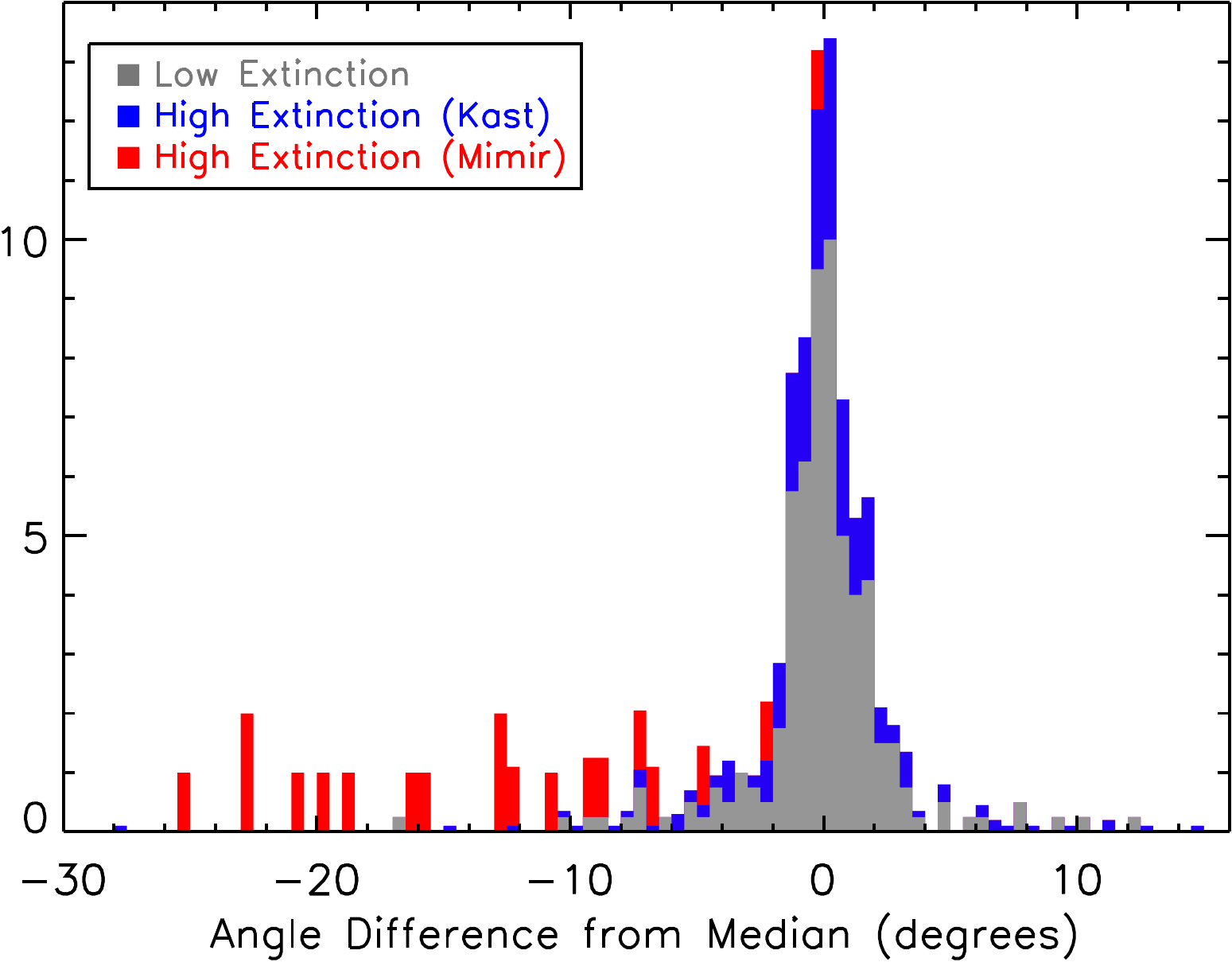}
  \caption{Stacked histograms of position angle differences with a
    $0\fdg5$
    binsize.  For each star, position angle differences were
    determined with respect to the median position angle measured for
    that star across the optical wavelength bins.  The numbers in each
    Kast and TURPOL data were normalized by the total number of
    wavelength bins so that the total area under the Kast and TURPOL
    histograms represents the total number of stars in the sample and
    is not biased by the different number of wavelength samples. The
    standard deviation of the low- and high-extinction distributions
    ($3\fdg7$
    and $4\fdg1$,
    respectively) are only slightly larger than typical angle
    uncertainties (e.g., Appendix~\ref{sec-angcal}).}
\label{fig-angles}
\end{figure}

The optical position angles in most of our sample are consistent with
constant angles across all measured wavelength bins.  Position angle
differences with respect to the median angle for each star are shown
in Figure~\ref{fig-angles}. For each star, position angle differences
were determined with respect to the median position angle measured for
that star across the optical wavelength bins. (Thus, the $H$-band
angle difference is not included in the median, but is plotted here as
the difference with respect to the median angle in the optical data.)
Due to the odd number of wavelength samples (11 and 5 for the Kast and
TURPOL data, respectively), one angle-difference sample is always
precisely zero when the median is subtracted; those data are not
plotted in Figure~\ref{fig-angles}. The numbers in each Kast and
TURPOL data bin were normalized by the total number of wavelength bins
so that the total area under the Kast and TURPOL histograms represents
the total number of stars in the sample and is not biased by the
different number of wavelength samples.

The standard deviations of the angle differences for the low- and
high-extinction distributions are $3\fdg7$ and $4\fdg1$, respectively.
These standard deviations are similar to typical uncertainties on the
angle measurements for each star in each wavelength bin,
$\sim$$2\arcdeg$--$5\arcdeg$.
In the low-extinction sample, only HD283772 shows a significant
rotation of position angles, from about $107\arcdeg$
to $92\arcdeg$
across $U$-
to $I$-band,
with uncertainties of $2\arcdeg$--$3\arcdeg$
per band.  Two stars in the high-extinction sample show significant
rotation over the wavelength span.  The angles for SWIW\,026 rotate
from about $20\arcdeg$--$25\arcdeg$
at $0.5\,\micron$
to $8\arcdeg$
at $1\,\micron$,
with typical uncertainties of $1\arcdeg$--$3\arcdeg$
per bin.  The angles for SWIW\,144 rotate from about $35\arcdeg$
at $0.5\,\micron$
to about $20\arcdeg$
at $1\,\micron$,
with typical uncertainties of $2\arcdeg$--$4\arcdeg$ per bin.

The last four columns of Table~\ref{tbl-serkowskifits} compare the
optical and infrared position angles for the 24 high extinction stars
with fits.  The mean optical position angles $\langle \theta \rangle$
are calculated by averaging the (variance-weighted) Stokes parameters
for all 11 bins. As an estimate of the uncertainty, we use the larger
of the standard deviation of the 11 angle measurements or the median
angle uncertainty of the 11 bins.  We make this conservative choice in
order to consider both statistical and systematic uncertainties in the
angle measurements.  Angles and uncertainties for the 20 stars with
$H$-band data are given the in the $\theta$(IR) and
$\sigma_\theta(\mathrm{IR})$ columns.  For the four stars without
$H$-band data, we use other measurements from the literature (see
references in Table) to estimate $\theta$(IR) and
$\sigma_\theta(\mathrm{IR})$.

The optical angles for these last four stars are in excellent
agreement with the existing IR measurements, within the uncertainties
of $1\arcdeg$--$2\arcdeg$.  However, the difference between the other
$H$-band and optical angles are larger (see the red bins in
Fig.~\ref{fig-angles}). Most significantly the optical and IR angles
for stars SWIW\,019, 057, 159, and 230 differ by more than three times
their uncertainties.

\subsection{Notes on Individual Stars} \label{sec-stars}

The goal of our work is to measure the polarization that the molecular
cloud imposes on unpolarized starlight.  Systematic biases will arise
in this measurement if any stars exhibit intrinsic polarization, such
as may occur with disks and other matter around young stars.  We check
for this possibility using observational tracers of circumstellar
matter, such as infrared (IR) excesses, or for indications that the
star is young, such as emission lines.

A physical association with the molecular cloud may be an indication
of young age and possible presence of circumstellar material. We first
check for such an association by comparing the distances of the cloud
to the distances of the stars in our sample. Distances to the stars
listed in Table~\ref{sourcetab} are obtained from the Gaia Data
Release 2 parallax survey
\citep{gaiadr2,bailer-jones2018}. \citet{galli2019} used Gaia DR2 and
radio VLBI astrometry to map the Taurus molecular cloud complex in
three dimensions; they find that B215 is the closest sub-structure in
the complex at a distance of $d=128.5\pm1.6$\,pc, while L1558 is the
most remote at $d=198.1\pm2.5$\,pc.

Of our low extinction sample (Table~\ref{turpol_source}), the star
BD+25\,698 ($d=121\pm1$\,pc) is located closer than B215 and an
additional 11 sources are at distance between the cloud extremes of
130 and 200\,pc (PPM- 93675, 93537, 93510, 93537, 93641, 93181, 93637,
93280; HD- 29334, 28170, 28975). None of these stars are found in
catalogs of young stellar objects in Taurus
\citep{herbigbell1988,rebull2010, rebull2011,herczeg2014}.  Hence the
low-extinction sample is unlikely to contain any sources with
significant circumstellar polarization.

Of the high-extinction sample (Table~\ref{sourcetab}), five stars lie
within the range 130 to 200\,pc (SWIW- 026, 040, 100, 121, 148). Two
of these stars are fairly distant from dense regions of the cloud and
thus unlikely to be YSOs associated with the cloud:
\begin{itemize}
\item SWIW026 ($d=158\pm1$\,pc) is projected closest to L1459 and
  ``Cluster 7'' of \citet{galli2019} at a distance of $130\pm1$\,pc.
\item SWIW100 ($d=164\pm3$\,pc) is located between L1495, probed by
  cluster 7 at $130\pm1$\,pc, and Heiles Cloud 2, probed by clusters
  14 and 15 at $142\pm2$\,pc and $138\pm2$\,pc, respectively.
\end{itemize}
While SWIW\,121 is close to L1531, it is most likely a Main Sequence
star, not a YSO. Based on our Lick/Kast spectra, we classify the star
as F0, with no observable emission lines. Similarly, no emission lines
were observed by \citet{coku1979} who classify the star as B2.

The remaining two stars are quite close to dense regions of the cloud,
increasing the likelihood that they are YSOs associated with the
cloud.
\begin{itemize}
\item SWIW040 ($d=131\pm1$\,pc) is projected closest to L1506, which
  is in line on the sky with B215 which is probed by ``cluster 2'',
  estimated to have a distance of $129\pm2$\,pc.
\item SWIW148 ($161\pm1$\,pc) is close to L1536 probed by cluster 16
  at $160\pm3$\,pc.
\end{itemize}
Additionally, SWIW\,040 and SWIW\,148 are T-Tauri stars
\citep{romano1975,kenyon1994,herbigbell1998} that exhibit emission
lines in our Lick/Kast spectra. Emission lines in SWIW\,040 were also
observed by \citep{luhman2009} and \citet{herczeg2014}, and
\citet{akimoto2019} attributes sporadic dimming of SWIW\,040 to
obscuration by a distorted circumstellar disk. Emission lines in
SWIW\,148 were also observed by \citet{herczeg2014}.

To further investigate the status of these stars, we used the
``Virtual Observatory Spectral Energy Density Analyzer'' (VOSA)
on-line tool \citep{bayo2008} to perform spectral energy distribution
(SED) fitting for the SWIW stars.  This Virtual Observatory tool uses
archival photometry spanning the UV (GALEX) to mid-infrared (WISE),
and allows fits based on many templates.  We used blackbody curves as
well as stellar models from the \citet{coelho2014} compilation.  The
derived spectral classes and extinctions are in good agreement with
our spectral classifications (Table~\ref{sourcetab}) and the
extinction values from \citet{shenoy2008}.  In this SED fitting,
SWIW\,040 and SWIW\,148 show clear evidence for infrared excess.
SWIW\,121 displays a marginal and wavelength-independent IR excess
possibly indicative of a debris disk. (Note that an IR excess does not
necessarily imply intrinsic polarization, as a circumstellar disk that
does not intercept the line of sight will cause IR excess, but not
polarization.)

\section{Discussion} \label{sec-discussion}

\subsection{Extinction vs.\
  \texorpdfstring{$\lambda_\mathrm{max}$}{Wavelength
    Peak}} \label{sec-discussion1}

As discussed in the introduction, radiative grain alignment theory
predicts a dependence of the alignment efficiency on the color of the
radiation field and the grain sizes, such that a grain \added{of
  radius $a$} is efficiently aligned when exposed to light of
wavelength $\lambda < 2a$ (unless the grain collision rate is
large---see below).  Therefore, as the color of the aligning radiation
field becomes redder (e.g., due to extinction) the size of the
smallest \emph{aligned} grain ($b_\mathrm{min} \approx \lambda/2$)
shifts to larger sizes, even as the small-grain end of the total
(aligned + unaligned) dust distribution remains fixed at
$a_\mathrm{min} < b_\mathrm{min}.$ Thus one expects
$\lambda_\mathrm{max}$ to increase with increasing $A_V$.

\begin{deluxetable*}{lcccccc}
\tablecaption{$\lambda_\mathrm{max}$-$A_V$ Relations
\label{lambdaavtable}}
\tablewidth{0pt}
\tablehead{
\colhead{} & \colhead{Number} & \colhead{$\alpha$}    & \colhead{$\sigma_\alpha$} & \colhead{$\beta$}         & \colhead{$\sigma_\beta$}   & \colhead{$A_V$} \\
\colhead{Data Set} & \colhead{of Stars} & \colhead{($\micron$)} & \colhead{($\micron$)}    & \colhead{($\micron$/mag)} & \colhead{($\micron$/mag)} & \colhead{(mag)} }%
\startdata
\citet{whittet2001}\tablenotemark{a} & 20 & \phs0.53\phn & 0.01\phn & 0.020\phn & 0.004\phn    & 0--4\phn \\
TURPOL, this work                    & 62 & \phs0.524    & 0.003    & 0.017\phn    & 0.001\phn    & 0--3\phn \\
TURPOL + \citet{whittet2001}         & 82 &\phs0.516    & 0.002    & 0.021\phn    & 0.001\phn    & 0--4\phn \\
all data in this work\tablenotemark{}  &   103         & \phs 0.508   & 0.016    & 0.0219       & 0.0007       & 0--10 \\
Lower Branch   & 96 &\phs0.513    & 0.002    & 0.019\phn    & 0.001\phn    & 0--6\phn \\
Upper Branch   & 7 & $-0.015$     & 0.10\phn &
0.17\phn\phn & 0.02\phn\phn & 0--10 \\
Upper Branch, wo/SWIW040   & 6 & \phs0.17\phn   & 0.11\phn & 0.12\phn\phn & 0.02\phn\phn & 0--10
\enddata
\tablecomments{Various subsets of data are fit to the linear relation
  $\lambda_\mathrm{max} = \alpha +\beta A_V$.  All fits are
  weighted by data uncertainties.  Data presented in this work are
  given in Tables~\ref{turpol_source}, \ref{sourcetab},
  \ref{tbl-serkowskifits-turpol}, and \ref{tbl-serkowskifits} and
  plotted in Figure~\ref{fig-data_models}.}

\tablenotetext{a}{Re-analyzed by \citet{bga2007}.}

\end{deluxetable*}

In order to assess the trends and differences in the data presented
here, we fit linear relations to various data subsets according to
\begin{equation}
  \lambda_\mathrm{max} = \alpha + \beta A_V
  \label{eq-linfit}
\end{equation}
(see also equation~[\ref{whittet_fit}]). The parameters, and parameter
uncertainties, of this linear model are given in
Table~\ref{lambdaavtable}.  Three sets of low-extinction data
(presented as separate rows in the table) include the sample of
\citet{whittet2001}, the TurPol sample from this work
(Table~\ref{turpol_source}), and the combined \citeauthor{whittet2001}
and TurPol samples. These data yield similar results, with
$\lambda_\mathrm{max}(\micron) \sim 0.52 + 0.02\,A_V$. A fit to the
complete data set varies only slightly from the low-extinction
data---the linear terms are the same (0.02\,\micron/mag) and the
constant term is $0.51\,\micron$.

While both terms for the ``all data'' case are similar to those of the
low-extinction samples, the uncertainty in the offset is larger in the
``all data'' case.  The cause of this much larger dispersion is a
number of outliers that systematically deviate from the low-extinction
fits around $A_V=3$--6\,mag (Figures
\ref{fig-data_models}\emph{a}--\emph{b}).

We split these points (SWIW 002, 019, 026, 040, 100, 101 and 144) into
an ``upper'' branch and leave the remaining points in a ``lower''
branch.  Separate linear fits to each branch are given in
Table~\ref{lambdaavtable}.
It is unlikely that the ``upper branch'' is only the result of noisy
data as an $F$-test finds that the addition of the ``upper branch'' is
justified at more than the 99\% probability level.  Additionally, the
fit to the ``upper branch'' is fairly robust against outliers---if
SWIW\,040, the star with the largest value of $\lambda_\mathrm{max}$
which is likely affected by instrinsic polarization from a
circumstellar disk of dust (\citealt{akimoto2019};
Section~\ref{sec-stars}), is excluded---the parameters of the upper
branch best fit are within 2$\sigma$ of the fits to the complete upper
branch (Table~\ref{lambdaavtable}). We also note that the seven
upper-branch stars are scattered throughout the cloud, not clustered
in any single region (Figure\,\ref{fig-map}), eliminating the
possibility that the upper-branch is a localized effect.


\subsection{Position Angles}

A detailed study of the angle differences, and their causes, is beyond
the scope of this work, but we note that the stars SWIW\,019, 026 and
144 show significant angle rotations with wavelength, well beyond the
measurement uncertainties.  These stars all lie on the ``upper
branch'' in the $\lambda_\mathrm{max}$-$A_V$ relation.  A similar
effect has been demonstrated for the star Elias 3-16 over the NIR
range by \citet{hough1988,hough2008}.  For that star, those authors
find that the position angle of the continuum polarization outside the
the $3.1\,\micron$ H$_2$O ice and the $4.7\,\micron$ CO ice lines are
$73^\circ\pm1.5^\circ$ and $73^\circ\pm2.3^\circ$, respectively.
Inside the two ice lines the position angles rotate to
$76^\circ\pm1.4^\circ$ in the H$_2$O ice line and
$86^\circ\pm4.2^\circ$ in the CO-ice feature.  The ice lines probe
only material at large extinction ($A_V>3.2$\,mag for H$_2$O ice and
$A_V>6.7$\,mag for CO ice \citep{whittet2007}), so the polarization
inside the lines probes magnetic fields at larger optical depths than
the continuum, where the weighting is more uniform along the line of
sight.

We argue that the observed position angle rotation on the ``upper
branch'' can be understood in a similar way. These lines of sight
likely probe both denser clumps as well as inter-clump gas. The grain
growth in the dense clumps however means that for the longest
wavelengths a relatively larger part of the polarization originates in
the clumps.  A systematic rotation in the magnetic field direction,
between the clump and inter-clump gas should therefore show up as a
rotation in the position angle with wavelength.  If this scenario is
correct, densely sampled multi-band (and ice line) polarimetry could
be used to probe the line-of-sight geometry of the ordered magnetic
field.

\subsection{A Grain Alignment Model}

The key point in our analysis is that the shape of polarization
spectrum (Eq.~\ref{serkowski}) is related to the dust-grain size
distribution. Any parameter that increases/decreases the size of the
smallest aligned grain, $b_\mathrm{min}$, shifts the peak in the
polarization spectrum to longer/shorter wavelengths. (A similar
relation holds between the spectrum and changes to the size of the
largest aligned grain, $b_\mathrm{max}$). This spectral shift is
parameterized by $\lambda_\mathrm{max}$ in Eq.~\ref{serkowski}.

The physical parameters of the aligned-grain size distribution are a
function of the underlying, total grain-size distribution and a
balance between processes that tend to align the grains and processes
that tend to dis-align the grains \citep{draine1998}.  Here, we
consider how the values of $b_\mathrm{min}$ and $b_\mathrm{max}$ vary
in different interstellar environments by applying RATs to align the
grains and gas-grain collisions to dis-align the grains. The inputs to
our model include
\begin{itemize}
\item A power-law form for the underlying grain-size distribution as
  given by \citet[MRN]{mathis1977}, but without the exponential
  extension proposed by later analyses
  \citep[e.g.,][]{kim1995,clayton2003}.
\item Gas-grain collision rates follow from the gas volume density $n$
  and temperature $T$.  Assuming that a grain will become disaligned
  once it has collided with its own mass in gas particles, the
  disalignment rate is proportional to \citep{hoang2015}
\begin{equation}
  R_\mathrm{dis}\propto \frac{n\times \sqrt{T_\mathrm{gas}}}{a}.
\end{equation}
Thus, smaller grains are more efficiently disaligned by gas-grain
collisions.  For all models presented here we set the temperature to a
constant value of 20\,K.  This simplification should not significantly
affect our results as the collision rate is only a weak function of
temperature ($R\sim T^{1/2}$).
\item Radiative torques are calculated using the local interstellar
  radiation field, as estimated by \citet{mathis1983}, and its
  extinction as a function of depth into the cloud. Radiative transfer
  is performed in only one dimension, using a plane-parallel slab
  geometry with the aforementioned single space density per model.
\end{itemize}
While the RATs are dependent upon the radiation field (including its
extinction $A_V$) and hence the \emph{column density}, the collisional
disalignment is dependent upon the gas \emph{volume density}.  Without
independent data on the volume density, we would have to assume a
physical path length in order to derive an average volume density from
the measured extinctions. Instead, we selected gas density values in
the range $n_\mathrm{gas}\sim10^2$--$10^4$\,cm$^{-3}$. We also note
that detailed three-dimensional modeling of radiative transfer and
grain alignment, which incorporates realistic cloud structures (e.g.,
self consistent density and temperature profiles or an explicit clumpy
structure with both clump and inter-clump gas in the same line of
sight), is beyond the scope of this work.

\begin{deluxetable*}{cDccrl}
  \tablecaption{Parameters of Grain Alignment Models\label{tbl-model}}
  \tablehead{ \colhead{Model} & \multicolumn2c{Gas Density} &
    \colhead{$b_\mathrm{max}$} &
    \colhead{$\lambda_\mathrm{max}(0)$\tablenotemark{a}}
    & \multicolumn2c{Figure}%
    \\
    \colhead{Number} & \multicolumn2c{($10^3$\,cm$^{-3}$)} &
    \colhead{($\micron$)} & \colhead{($\micron$)}
    & \multicolumn2c{Legend\tablenotemark{b}}%
  }%
  \decimals 
  \startdata
  1  &  0.1  &  0.3   & 0.49   &   solid, &black \\
  2  &  10   &  0.3   & 0.55   &   solid, &red \\
  3  &  10   &  0.5   & 0.55   &  dotted, &red \\
  4  &  40   &  0.3   & 0.65   &   solid, &blue \\
  5  &  40   &  0.6   & 0.65   &  dashed, &blue \\
  6  &  40   &  0.7   & 0.65   &   solid, &green \\
  \enddata
  \tablecomments{All models have a gas temperature of 20\,K.}
  \tablenotetext{a}{Limit as extinction approaches zero;
    $b_\mathrm{min} = a_\mathrm{min}$.}  \tablenotetext{b}{Line color
    and type used in Figures~\ref{fig-data_models}\emph{c}--\emph{d}.}
\end{deluxetable*}

Model results are shown in Figure~\ref{fig-data_models}\emph{c} and
compared to the data in Figure~\ref{fig-data_models}\emph{d}.  The
primary features of the model plots are (1) at low extinction,
$\lambda_\mathrm{max}$ has a minimum that varies with volume density
but is independent of $b_\mathrm{max}$; (2) for models with the same
volume density $\lambda_\mathrm{max}$ reaches a maximum that increases
as $b_\mathrm{max}$ increases.  For example, models 2 and 3, both with
a density of $10^4$\,cm$^{-3}$, converge at
$\lambda_\mathrm{max}=0.55\,\micron$. The divergence of these two
models above an extinction of 6\,mag is due to the different values of
$b_\mathrm{max}$.  Similarly, models 4, 5, and 6, all with a density
of $4\times10^4$\,cm$^{-3}$, converge at
$\lambda_\mathrm{max}=0.65\,\micron$ but diverge at moderate
extinction levels, reaching larger values of $\lambda_\mathrm{max}$ as
$b_\mathrm{max}$ increases from 0.3 to $0.7\micron$.

\subsection{Dispersion in
  \texorpdfstring{$\lambda_\mathrm{max}(A_V)$}{Wavelength
    Peak}} \label{sec-dispersion}

At low extinctions ($<$4\,mag; particularly near $A_V=1.5$--2.5\,mag)
the expanded sample shows an enhanced dispersion of
$\lambda_\mathrm{max}$ compared to the \citet{whittet2001} sample.
Thus, the size distributions of aligned grains must differ along these
lines of sight, despite having similar extinction levels (also see
\citealt{wang2017}).

For a given extinction level, one expects the radiation fields to be
similar, unless the extinction from the observer through the target
cloud region is not well-correlated with the extinction to the cloud
grains from the illuminating source.  However, stars with obvious
anomalous sightlines were pre-screened \citep[e.g.,][]{bga2007} and
removed from the list of targets in Table~\ref{turpol_source}.

Another possibility for this dispersion is grain growth. The ratio of
total-to-selective extinction, $R_V$, is generally correlated with
grain size \citep{nozawa2016,cardelli1988}. Estimates of $R_V$ for the
low-extinction line-of-sight sample (Table~\ref{turpol_source}) are
found using spectral classifications \citep{wright2003} and optical
and NIR photometry \citep{tycho2000,skrutskie2006}.
For the extinction ranges $A_V=0.5$--1.5, 1.5--2.5 and 2.5--3.5\,mag,
$R_V$ has average values of $3.20\pm0.08$, $3.27\pm0.07$ and
$3.66\pm0.27$, respectively (data in
Table~\ref{turpol_source}). Restricting the middle range of $A_V$ to
only stars with a large deviation
($\delta\lambda_\mathrm{max}>+0.05\,\micron$) from the linear
$\lambda_\mathrm{max}$-$A_V$ relation, the average is unchanged at
$\langle R_V\rangle=3.21\pm0.19$.  These small variations in $R_V$
yield no strong evidence for grain growth within or between these
three extinction ranges.

\begin{figure*}
  \plottwo{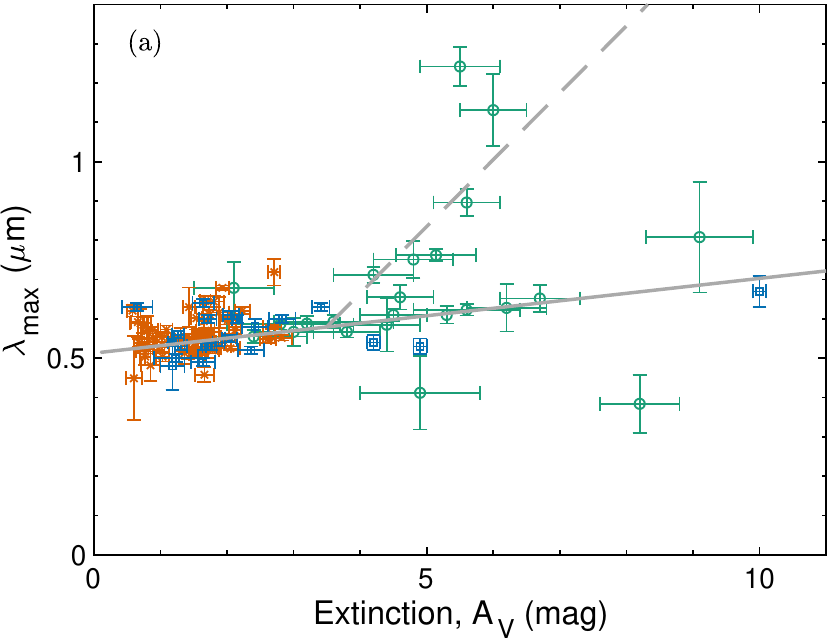}{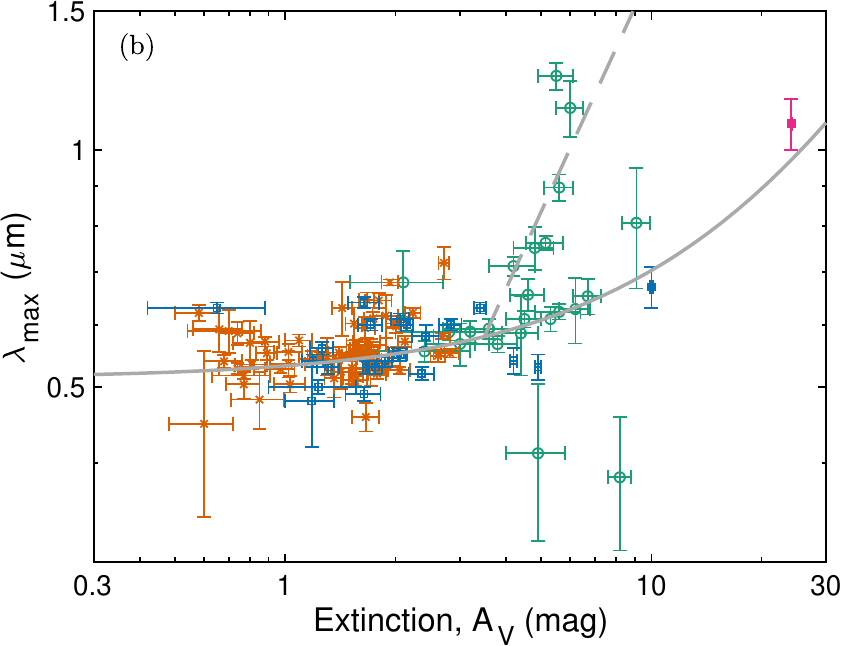}\\
  \hskip 1in\\
  \plottwo{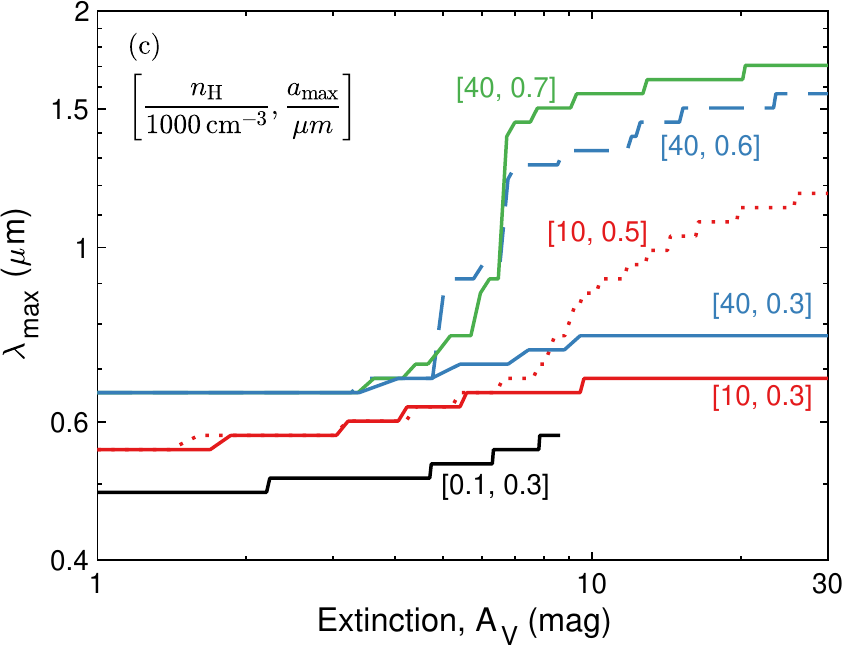}{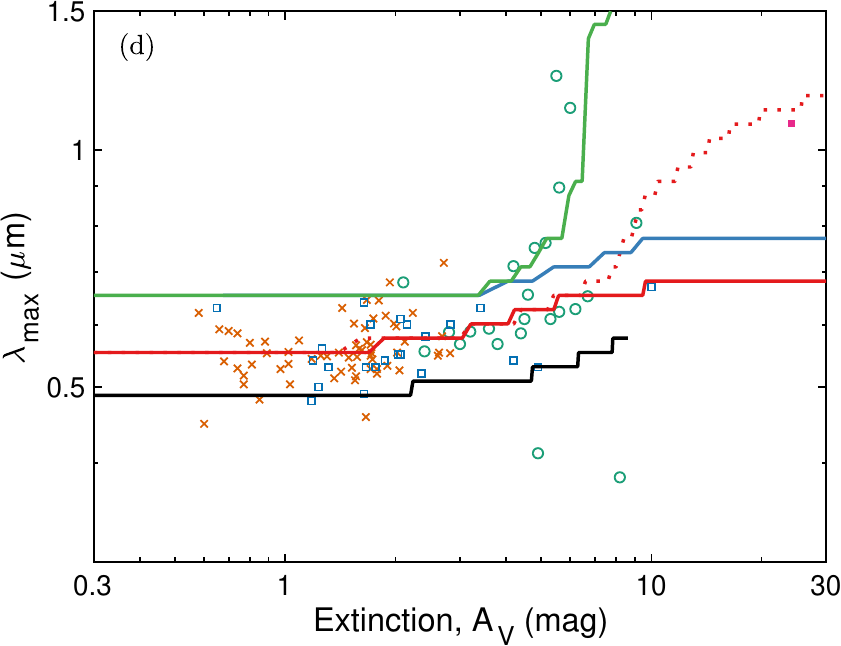}
  \caption{Panels (\emph{a}) and (\emph{b}) show the relation between
    the wavelength peak of the polarization curve
    ($\lambda_\mathrm{max}$; equation~[\ref{serkowski}]) and optical
    extinction ($A_V$) towards stars behind Taurus. Data are from both
    this work (Kast/Lick = green and TurPol/NOT = orange) and
    \citeauthor{whittet2001}\ (\citeyear{whittet2001}; blue).  Panel
    (\emph{a}) is plotted using linear axes while panel (\emph{b}) is
    plotted logarithmically and also includes the data point for Elias
    3-16 (\citealt{hough1988,murakawa2000}; magenta).  The solid and
    dashed lines show the best linear fits to the ``upper'' and
    ``lower'' branches (Table~\ref{lambdaavtable}), respectively.
    (\emph{c})~Model fits based on RAT alignment are shown for a
    number of model parameters.  Lines are labeled with the volume
    density (in units of 1000\,cm$^{-3}$) and the maximum grain size
    ($\micron$) in each model. In order to show the key model
    differences, the plot shows a slightly different range of $A_V$
    and $\lambda_\mathrm{max}$ from panels \emph{a}, \emph{b}, and
    \emph{d}.  Panel (\emph{d}) compares several of the models in
    panel \emph{c} (using the same color scheme) with the data in
    panels \emph{a} and \emph{b}.  The error bars on the data are
    removed for clarity only.}
  \label{fig-data_models}
\end{figure*}

Thus the increased dispersion of $\lambda_\mathrm{max}$ for the low
extinction sightlines is unlikely to be the result of either changes
in the radiation field (which should be fully determined by $A_V$) or
grain growth (which would change $R_V$).  However, these data do fall
within the range for which our modeled values of
$\lambda_\mathrm{max}$ are most efficiently varied by changing the
volume density---the data are effectively bounded by models with
densities (0.1 to 40)$\times10^3$\,cm$^{-3}$
(Figure~\ref{fig-data_models}\emph{d}). Thus the dispersion is most
likely a result of an increased gas-grain collision rate which causes
an increase in $b_\mathrm{min}$. Given the low average density in the
cloud ($\sim 1\times10^3$\,cm$^{-3}$; \citealt{BlitzWilliams1997}) the
areas of higher collision may result from higher-density clumps along
these lines of sight, especially at $A_V<0.8$\,mag where the low
extinction would indicate lower volume densities than the
$A_V=1.5$--2.5\,mag sample.

\subsection{Comparison of Models Data}

Between $A_V\sim1$ and 6\,mag, model 2 (red line in
Figure~\ref{fig-data_models}\emph{d}) reproduces the observed linear
relationship in the lower-branch data (grey, solid, curve in
Figures~\ref{fig-data_models}\emph{a}--\emph{b}), before flattening
out beyond $A_V\sim10$\,mag.  This flat part of the curve results from
a lack of grains beyond the upper-size cut-off of $0.3\,\micron$ that
satisfy the RAT alignment condition ($a>2\lambda$) for the remaining,
reddened, radiation field.
To reach the measured values for Elias 3-16 ($A_V=24.1$\,mag,
$\lambda_\mathrm{max}=1.08\,\micron$) we increased the largest grain
size to $b_\mathrm{max}=0.5\,\micron$ (model 3).  This new curve only
deviates from the baseline model beyond $A_V\approx 6$\,mag,
confirming that these values of $b_\mathrm{max}$ do not strongly
influence the results at low extinction.

To reproduce the upper branch data, in which
$\lambda_\mathrm{max}>0.8\,\micron$ above 6\,mag, it was necessary to
increase both the size of largest grains in the model and the gas
density. The resulting increase in $b_\mathrm{min}$ due to collisions,
together with the larger value of $b_\mathrm{max}$, increases the
average size of aligned grains, and hence increases
$\lambda_\mathrm{max}$ significantly.  Models 4--6 yield equivalent
$\lambda_\mathrm{max}$ at low extinction because they all use the same
volume density ($4\times 10^4$\,cm$^{-3}$), but increase to larger
values of $\lambda_\mathrm{max}$ as $b_\mathrm{max}$ increases. Model
7 with $b_\mathrm{max}=0.7\,\micron$ yields the best match to the
upper branch data (Figure~\ref{fig-data_models}\emph{d}).

Grain growth in dense clouds is thought to be dominated by grain
coagulation \citep{ormel2009}, which is a collisional process.  This
is especially so for the silicates \citep{hirashita2014}, which are
responsible for polarization.  The upper-branch models require both
enhanced gas densities and increases in
$a_\mathrm{max}$($=b_\mathrm{max}$), neither change alone is
sufficient. Such a correlation of increased gas density and larger
grain sizes is expected for models of grain growth through
coagulation.  Therefore, the bifurcation of the $\lambda_\mathrm{max}$
vs.\ $A_V$ relationship into two branches may indicate the presence of
a clumpy volume density structure, and possibly fractal cloud
structures \citep{falgarone1991b,falgarone1996} where, even for
high-$A_V$ regions, only some lines of sight probe dense material with
significant grain growth. Note that the strong bifurcation between the
two branches suggests that any given sight line is dominated by either
strong grain growth (the upper branch with
$a_\mathrm{max}=0.7\,\micron$) or only moderate growth (the lower
branch with $a_\mathrm{max}<0.5\,\micron$). Otherwise one would expect
more points to fall between the two branches.  Polarization spectra of
additional stars at high extinction are needed to sample the region
between the branches.

Based on scattering theory one expects that both
$\lambda_\mathrm{max}$ and $R_V$ depend on grain size and, thus, that
the two parameters are correlated. We used Gaia photometric data to
estimate the total-to-selective extinction ratio towards the SWIW
sample using two methods: 1) using the equation
$R_V=1.1\cdot E_{V-K}/E_{B-V}$ \citep{whittet1978}; and 2) by fitting
a second order polynomial to $E_{\lambda-V}/E_{B-V}$ as a function of
$1/\lambda$ and finding
$R_V=\lim_{\lambda \to \infty} E_{\lambda-V}/E_{B-V}$
\citep[c.f.,][]{whittet2003}. The results from the two methods, shown
in Table~\ref{sourcetab}, generally agree well.  (We found that fits
based on color-transformed SDSS or Pan-STARRS
\citep[e.g.][]{jester2005} data tended to yield extreme, and often
un-physical results and have, therefore, used the AAVSO ($B$,$V$) data
where available.)
  
\begin{figure}
  \centering 
  \plotone{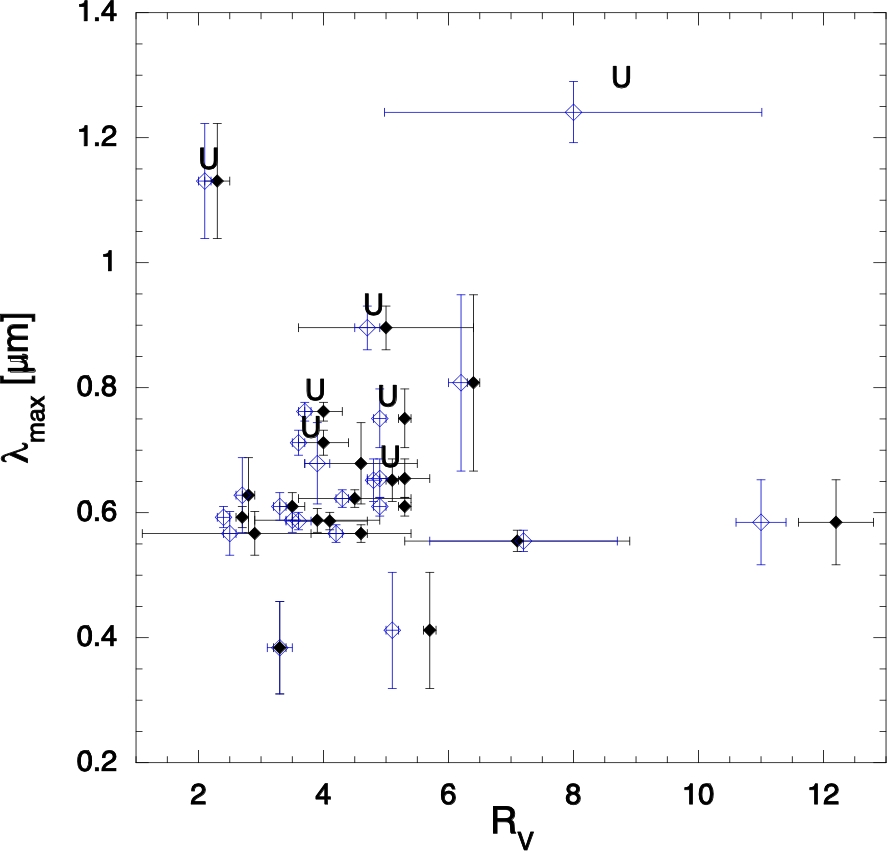}
  \caption{Extinction ratio and $\lambda_\mathrm{max}$. The
    total-to-selective extinction ratio ($R_V$) for the SWIW sample is
    plotted with the measured values of $\lambda_\mathrm{max}$. The
    blue circles represent values derived from the relation
    $R_V = 1.1 \cdot E_{V-K}/E_{B-V}$ \citep{whittet1978}, while the
    red triangles represent values derived from
    $R_V=\lim_{\lambda \to \infty} E_{\lambda-V}/E_{B-V}$
    \citep[c.f.][]{whittet2003}.  Stars on the upper branch of
    Figure~\ref{fig-data_models} are labeled as ``U'', while lower
    branch stars remain unlabeled.}
  \label{rv_vs_lmax_fig}
\end{figure}

Values for $\lambda_\mathrm{max}$ and $R_V$ for the high-density
sample are compared in Figure~\ref{rv_vs_lmax_fig}. While a
significant number of outliers are apparent in this plot, a weak
correlation is apparent in the central part
($\lambda_\mathrm{max}=0.55$--$0.90\,\micron$, $R_V=2$--6). The
upper-branch stars (labeled ``U'' in the figure) show both larger
$\lambda_\mathrm{max}$ and slightly larger $R_V$ from the lower-branch
(unlabeled) stars.  Given the many uncertainties associated with
calculating $R_V$ for these stars (e.g., spectral class assignments
affecting the intrinsic colors, emission lines affecting the observed
colors), as well as small-number statistics it is not surprising that
the apparent correlation between $\lambda_\mathrm{max}$ and $R_V$ is
small in our sample. Because $\lambda_\mathrm{max}$ is not affected by
the above stellar and photometric uncertainties, we expect that it is
a more direct measure of the average grain size along the line of
sight than is $R_V$.

\section{Conclusions}\label{Conclusion_sec}

We have acquired and analyzed an expanded sample of multi-band
photo-polarimetry at moderate extinctions (62 stars), and visible
spectro-polarimetry of high-extinction lines of sight (25 stars),
through the Taurus cloud complex.  To the visible spectropolarimetry
data, we add NIR $H$-band polarimetry in order to better constrain
polarization vs.\ wavelength fits to the Serkowski relation parameters
($p_\mathrm{max}$, $\lambda_\mathrm{max}$, $K$).

We confirm the previously established \citep{whittet2001,bga2007}
\replaced{linear relationship}{correlation} between
$\lambda_\mathrm{max}$ and $A_V$ for extinctions up to
$A_V\approx4$\,mag.
Beyond $\sim$4\,mag, the $\lambda_\mathrm{max}$ vs.\ $A_V$
relationship bifurcates, with part of the sample continuing the
previously observed relation (a ``lower branch'') while another part
of the sample has a significantly steeper dependence of
$\lambda_\mathrm{max}$ on $A_V$ (an ``upper branch'').

Using RAT modeling of the grain alignment and radiative transfer in
the cloud, we
find that the lower branch data are well modeled by RAT alignment of
grains with a fixed size distribution, illuminated by an increasingly
reddened diffuse interstellar radiation field, and a constant level of
gas-grain collisional disalignment. For lines of sight having
$A_V=1.5$--2.5\,mag and enhanced $\lambda_\mathrm{max}$ values,
increased collisional disalignment alone can explain the observed
behavior, consistent with the lack of an increase in the
total-to-selective extinction ($R_V$) for these lines of sight.

For the upper branch, both grain growth and increased collisional
disalignment of the smallest grains are required to match the
observations \citep[c.f.,][]{whittet2003}.  That the grain growth on
the upper branch is associated with enhanced volume density is
consistent with the expectation that grain growth through coagulation
is a collisional process, and therefore will proceed faster in denser
material.

Our results indicate that multi-band polarization can be used as a new
tool to probe grain growth in molecular clouds, independent of
assumptions about grain temperature and emissivity required for the
combination of near- and far-infrared data \citep{ysard2013}.

\acknowledgements 

The authors would like to thank Thomas Lowe and other members of the
Lick Observatory staff for assistance with observations.  We would
also like to thank Ryan Chornock for advice on data analysis with the
Kast spectropolarimeter, Dr. Miguel Charcos-Llorens for assistance
with spectral classification, and Ilija Medan for a careful reading of
the manuscript and many useful comments.

B-G A. gratefully acknowledges a grant from the National Science
Foundation (1109469 and 1715867) which made this work possible.
J. E. V. acknowledges support from NASA under award No.\ SOF 05-0038.
T. H. acknowledges the support of the Basic Science Research Program
of the National Research Foundation of Korea (NRF), funded by the
Ministry of Education (2017R1D1A1B03035359).

Mimir observations in 2011 September were performed by Ren Cashman and
those in 2012 by Brian Taylor.  This research was conducted in part
using the Mimir instrument, jointly developed at Boston University and
Lowell Observatory and supported by NASA, NSF, and the W. M. Keck
Foundation.  This effort was made possible by grants AST 06-07500, AST
09-07790, AST 14-12269 and AST 18-14531 from NSF/MPS to Boston
University and by grants of observing time from the Boston University
-- Lowell Observatory partnership.

This research has made use of the ``Aladin sky atlas'' (developed at
CDS, Strasbourg Observatory, France) and NASA's SkyView facility
(http://skyview.gsfc.nasa.gov; located at NASA Goddard Space Flight
Center).


\facilities{NOT (TurPol), Perkins (Mimir), Shane (Kast Double
  spectrograph), Planck (HFI), Gaia}

\appendix

\section{Data Reduction for Optical
  Spectropolarimetry} \label{sec-panalysis} 

This appendix discusses details of the polarization analyses performed
on the data collected with the Kast spectropolarimeter on the 3\,m
Shane telescope of Lick Observatory (Section~\ref{sec-kast};
\citealt{miller1988}).

\subsection{The Polarization Signal}

The spectral images for all stars and standards were flat-fielded at
each of the eight HWP angles.  Spectra of the two orthogonal
polarizations (the ordinary, $O$, and extraordinary, $E$, rays) were
separately wavelength-calibrated and extracted using standard
IRAF\footnote{IRAF is distributed by the National Optical Astronomy
  Observatories, which are operated by the Association of Universities
  for Research in Astronomy, Inc., under cooperative agreement with
  the National Science Foundation. (\url{http://iraf.noao.edu/})}
routines in the APEXTRACT package.

Given the extraordinary $E(\lambda,\theta)$ and ordinary
$O(\lambda,\theta)$ spectral signals, at HWP angle $\theta$ and
wavelength $\lambda$, we defined the difference and sum (Stokes $I$)
signals as
\begin{eqnarray}
d(\lambda,\theta) & \equiv & O(\lambda,\theta)-E(\lambda,\theta) \label{eq-dif} \\
\mathrm{and} \quad 
I(\lambda,\theta) & \equiv & O(\lambda,\theta)+E(\lambda,\theta) \label{eq-sum}.
\end{eqnarray}
The polarization signal is then
\begin{equation}
  S(\lambda,\theta) = \frac{d(\lambda,\theta)}{I(\lambda,\theta)}.
\end{equation}
For an ideal HWP, this signal has the form
\begin{eqnarray}
  S(\lambda,\theta,\delta) & = & a(\lambda) + p(\lambda)\,\cos4[\theta-\delta(\lambda)] \\
 & = & a(\lambda) + q(\lambda)\,\cos4\theta + u(\lambda)\,\sin4\theta 
       \label{eq-stokes}
\end{eqnarray}
where (dropping the $\lambda$-dependence for simplicity) $p$ is the
polarized fraction, $\delta$ is the phase angle of the measured
polarization, and $q$ and $u$ are the reduced-Stokes parameters,
$q\equiv Q/I=p\,\cos 4\delta$ and $u\equiv U/I = p\,\sin4\delta$. The
offset factor $a$ accounts for gain differences (between the $E$- and
$O$-beams of the Wollaston prism) that have not been completely
corrected by the flat-fielding analysis step
\citep[e.g.,][]{patat2006}.  Note that $I$ is the total intensity
Stokes parameter and is, ideally, equivalent to that in equation
(\ref{eq-sum}) and expected to have no $\theta$-dependence. The phase
angle and polarization are related to the Stokes parameters via
\begin{eqnarray}
  p & = & \left(q^2 + u^2\right)^{1/2} \label {eq-p}\\
  \mathrm{and} \quad 
  2\delta & = &  \frac{1}{2} \arctan\left(\frac{u}{q}\right), 
                \label{eq-delta}
\end{eqnarray}
where we have again dropped the $\lambda$-dependence on all four
parameters.  The polarization angle in space, $2\delta$, is related to
the stellar polarization position angle $\theta$ projected onto the
sky and several instrument parameters (see Appendix~\ref{sec-angcal}).

To improve the signal-to-noise on individual measurements, the spectra
for the sum $I(\lambda, \theta)$ and difference $d(\lambda,\theta)$
signals in the 0.460--$1.010\,\micron$ range were averaged into 11
bins, each $0.050\,\micron$ wide, with equal weighting applied to all
spectral pixels in a bin.  Nominally, the uncertainty in each bin
would be the standard deviation of the mean.  However, as described in
Section~\ref{sec-obs}, the spectral resolution of our observations was
typically 3--5 on-chip pixels FWHM ($\approx0.0020\,\micron$), so that
the individual pixels in an averaged bin were not statistically
independent.  To correct for this, the uncertainty on the difference
signal in each $0.050\,\micron$ bin was set to twice the standard
deviation of the mean.  The uncertainty on the sum signal at every
$\theta$ was taken to be the standard deviation of the measurements
across all eight HWP angles.  Since the intensity $I(\lambda,\theta)$
was nominally independent of HWP angle $\theta$, its repeatability was
used as an estimate of the measurement uncertainty.  This variation is
most likely the result of a time-variable sky transmission which we
have not attempted to remove here \citep[e.g.,][]{clemens2012b}. The
uncertainties on the sum and difference signal were propagated into
those for the polarization signal $S(\theta,\lambda)$.

Empirically, the different uncertainties for $I$ and $d$ at each HWP
angle in a given wavelength bin were comparable.  Occasionally some
uncertainty values did differ significantly, which was not unexpected
given the large data set. To avoid over- or under-weighting these data
in the fits, we assign uniform uncertainties $\sigma_u(\lambda)$ to
the polarization signals $S(\lambda,\theta)$ within each wavelength
bin.  Within each wavelength bin, the uniform uncertainties are given
by the median uncertainty across the HWP angles $\theta_i$ in that
bin, such that
$\sigma(\lambda,\theta_i) = \sigma_u(\lambda)
=\mathrm{median}[\sigma(\lambda,\theta_i)]$.

\subsection{Polarization Fits} \label{sec-pfits}

Equation (\ref{eq-stokes}) describes a set of coupled linear equations
whose solution is found by performing a standard linear
regression. That is, we wish to solve the coupled equations
\begin{equation}
  \boldsymbol{S} = \mathbf{A}\boldsymbol{x},
  \label{eq-matrix1}
\end{equation}
where the data vector is
\begin{equation}
  \boldsymbol{S}(\lambda) = \left[ \begin{array}{c} 
                                     S(\theta_1,\lambda)  \\
                                     \vdots \\
                                     S(\theta_n,\lambda)
                                   \end{array} \right]
                                 \label{eq-datamatrix1}
\end{equation}
and $n$ is the number of HWP angles (typically $n=8$). The parameter
matrix with the fit Stokes parameters is
\begin{equation}
  \boldsymbol{x}(\lambda) = \left[ \begin{array}{c} 
                                     a(\lambda) \\
                                     q(\lambda) \\
                                     u(\lambda)
                                   \end{array} \right],
                                 \label{eq-datamatrix2}
\end{equation}

\begin{equation}
  \mathbf{A} = \left[ \begin{array}{ccc} 
                        1 & \cos 4\theta_1 & \sin 4\theta_1 \\
                        \vdots & \vdots & \vdots \\
                        1 & \cos 4\theta_n & \sin 4\theta_n \\
                      \end{array} \right].
                    \label{eq-amatrix1}
\end{equation}
The least-squares solution to equation~(\ref{eq-matrix1}) is
\begin{equation}
  \tilde{\boldsymbol{x}} = \left(\mathbf{A}\!^\mathrm{T}\,\mathbf{\Sigma}^2\,\mathbf{A} \right)^{-1}\mathbf{A}\!^\mathrm{T}\,\mathbf{\Sigma^2}\boldsymbol{S}.
  \label{eq-amatrix2}
\end{equation}
where the matrix of inverse-variances is diagonal with elements
$\Sigma^2_{ii}=\left[1/\sigma(\theta_i,\lambda)\right]^2$.

Parameter uncertainties follow from the diagonal elements of the
covariance matrix such that
\begin{equation}
  \left(\mathbf{A}\!^\mathrm{T}\,\mathbf{\Sigma}^2\,\mathbf{A}
  \right)^{-1} =
  \sigma^2(\lambda)\left[\mathbf{A}\!^\mathrm{T}\mathbf{A} \right]^{-1}
  = 
  \left[ \begin{array}{ccc}
           \sigma_{aa}^2  & \sigma_{aq}^2 & \sigma_{au}^2 \\
           \sigma_{aq}^2  & \sigma_{qq}^2 & \sigma_{qu}^2 \\
           \sigma_{au}^2  & \sigma_{qu}^2 & \sigma_{uu}^2 \\
         \end{array} 
       \right],
       \label{eq-amatrix3}
\end{equation}
where the first equality holds when the uncertainties are the same for
all HWP angles.  For the typical case with eight equally spaced HWP
angles from zero to 157.5, degrees all off-diagonal terms reduce to
zero and the diagonal terms simplify to
$\sigma_{aa}(\lambda) = \sigma(\lambda)/\sqrt{8}$ and
$\sigma_{qq}(\lambda) = \sigma_{uu}(\lambda) = \sigma(\lambda)/2$.


The least-squared solutions also return reduced-$\chi^2$
goodness-of-fit parameters for each wavelength bin for each star.  Of
the 275 fitted data points (11 wavelengths $\times$ 25 stars) the
returned $\chi^2$ values have a log-normal distribution with median
0.55 and standard deviation 0.7.

The polarization amplitude, polarization phase angle, and their
respective uncertainties follow from equations (\ref{eq-p}) and
(\ref{eq-delta}).  Note that some measurements did not include all
eight HWP angles.  For those objects the off-diagonal covariance terms
cannot necessarily be ignored and must be included when calculating
the amplitude and phase uncertainties.  However, in practice we found
that the covariance terms were much smaller than the diagonal terms
for our particular dataset and so did not include them in the
calculation.

\section{Polarization Calibration for Optical
  Spectropolarimetry} \label{sec-pcal}

\subsection{Polarization Standards} \label{sec-highp}

\begin{figure}
  \centering
  \includegraphics[height=5.6cm]{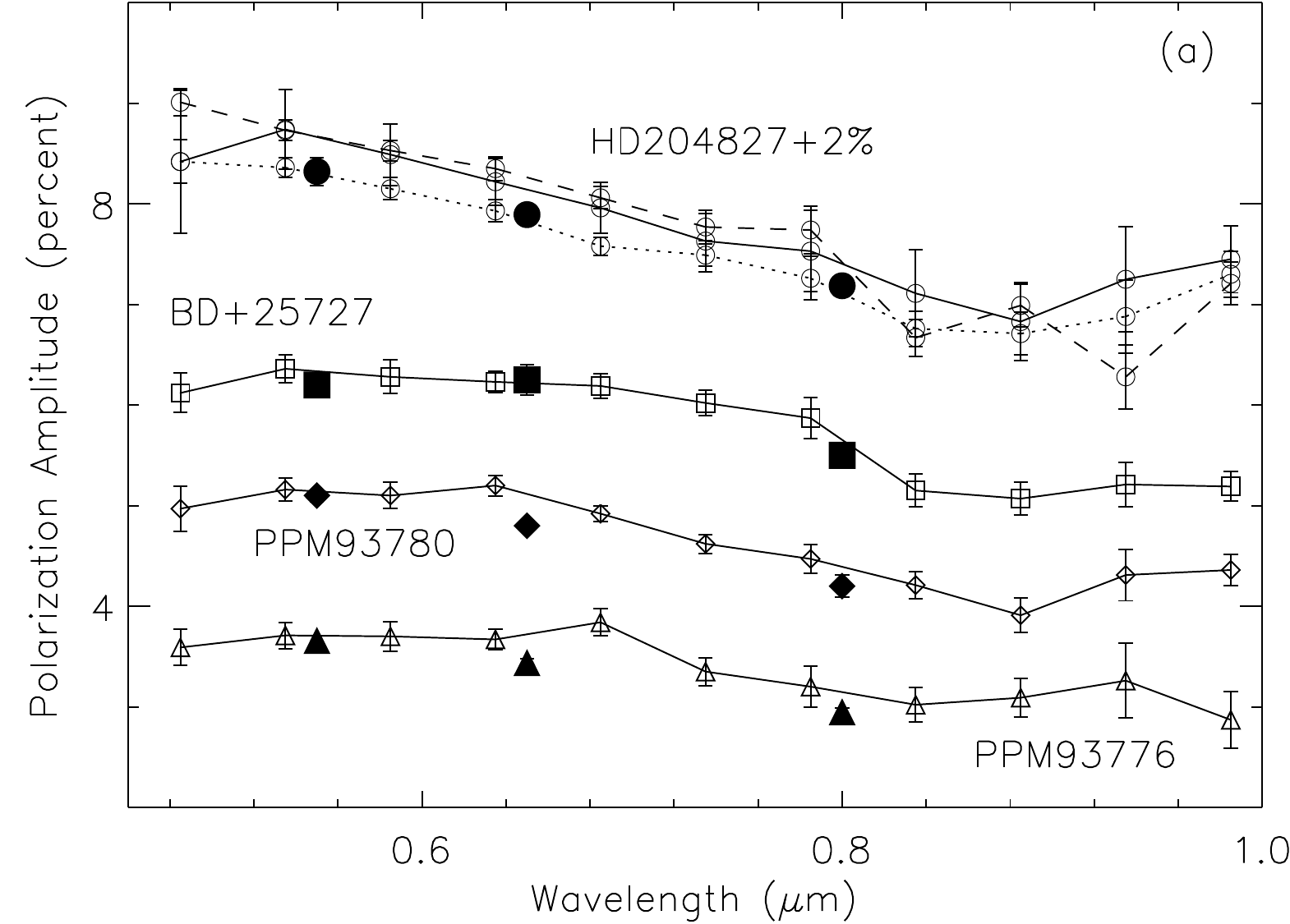}
  \hfil
  \includegraphics[height=5.6cm]{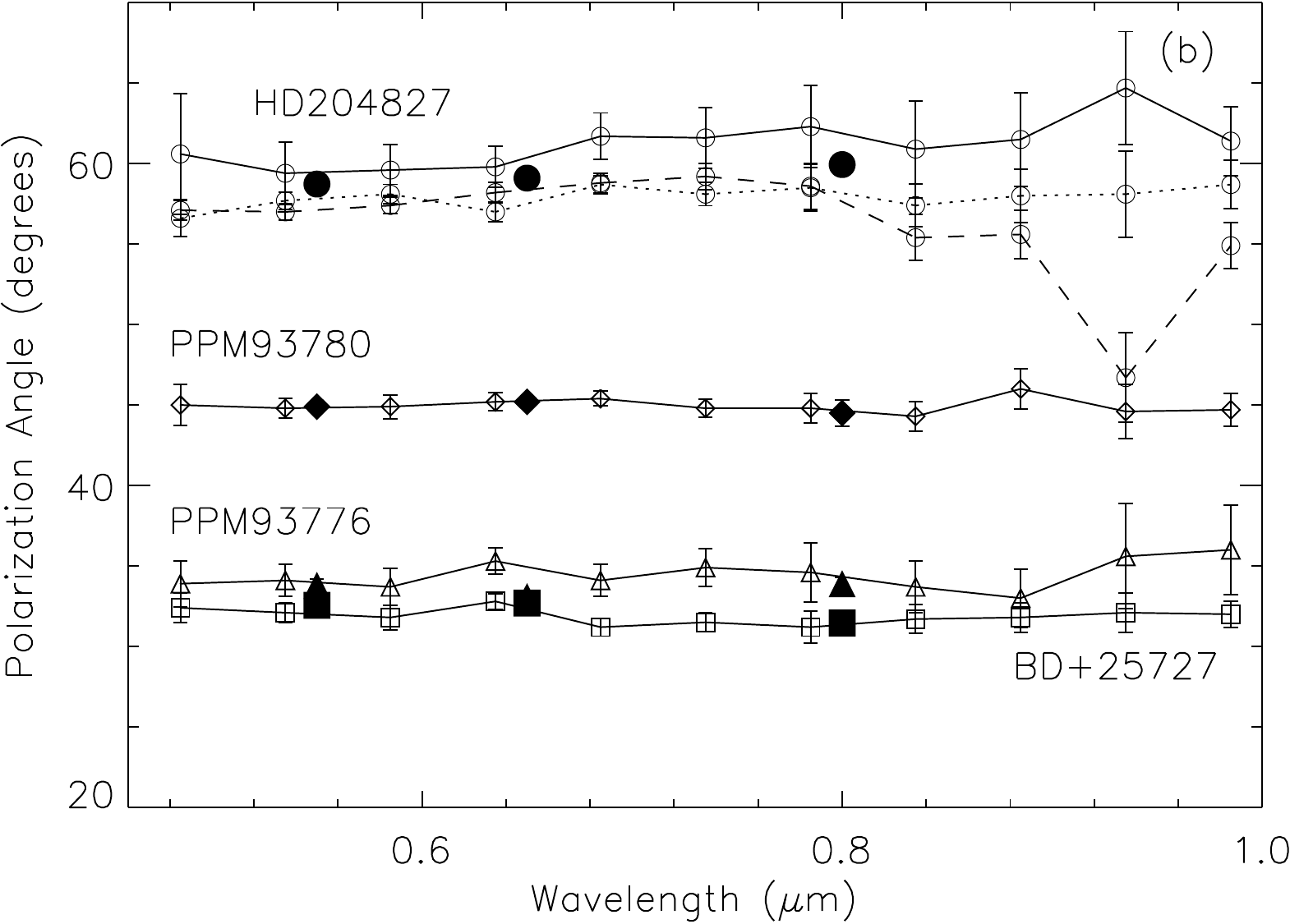}
  \caption{Polarization amplitudes (\emph{a}) and angles (\emph{b})
    measured for standard high-polarization stars (see
    Table~\ref{tbl-polstds}).  All three separate measurements for
    HD\,204827 are shown as open circles with different observations
    indicated by different line-types. Error-bars represent only the
    statistical uncertainties in the polarization fits and do not
    include any estimate of systematic uncertainties. Large solid
    symbols show TurPol measurements of all four stars through
    broadband $V$ (0.55\,\micron), $R$ (0.65\,\micron), and $I$
    (0.80\,\micron) filters (see text for references and band
    definitions).  Uncertainties for data without error-bars are
    smaller than the plotted symbols.  For clarity, the polarization
    amplitude data for HD\,204827 have been shifted up by 2\%.}
  \label{fig-highp}
\end{figure}

Table~\ref{tbl-polstds} lists observations of several known
high-polarization
stars\footnote{\url{http://www.not.iac.es/instruments/turpol/std/hpstd.html}}
\citep{schmidt1992,turnshek1990}.  We used these observations as a
check against systematic errors introduced via the observations with
the Kast spectropolarimeter or the data analysis process, and to
calibrate the angular position of the Kast instrument.  Two stars were
used throughout the three nights of observations, with BD+25727
observed only on the first night and HD\,204827 on all three nights.
Polarization results for all 11 passbands are given in
Figure~\ref{fig-highp} and Table~\ref{tbl-polstds}. For comparison, we
also plot the broadband measurements from the literature for
HD\,204827 \citep{schmidt1992}, results from the Nordic Optical
telescope for
BD+25727\footnote{\url{http://www.not.iac.es/instruments/turpol/std/bd25727\_note.html}},
and the results of our TurPol observations for the stars PPM\,93776
and PPM\,93780.

\begin{deluxetable}{lccccccccccc}
\rotate
  \tablewidth{0pt} \tablecaption{Standard High-polarization Stars
    observed using
    Kast \label{tbl-polstds}} \label{tbl-8}
    \tablehead{%
    \colhead{Object} & \colhead{Band} & \colhead{Center
      Wavelength\tablenotemark{a}} & \colhead{$q$} &
    \colhead{$\sigma_q$} & \colhead{$u$} & \colhead{$\sigma_u$} &
    \colhead{$p$} & \colhead{$\sigma_p$} &
    \colhead{$\theta$\tablenotemark{b}} & \colhead{$\sigma_\theta$} &
    \colhead{$\chi_r^2$}
    \\
    \colhead{} & \colhead{} & \colhead{($\micron$)} & \colhead{(\%)} &
    \colhead{(\%)} & \colhead{(\%)} & \colhead{(\%)} & \colhead{(\%)}
    & \colhead{(\%)} & \colhead{(deg.)} & \colhead{(deg.)} &
    \colhead{} }

\startdata
BD\,+25727 & Narrow   &  0.485 & \phs2.44 &  0.20 &   5.61 &  0.20 &  6.12 &  0.20 &   33.2 &  0.9 & \phn2.00 \\
BD\,+25727 &  Narrow  &  0.535 & \phs2.60 &  0.14 &   5.81 &  0.14 &  6.36 &  0.14 &   32.9 &  0.6 &  \phn0.45 \\
\vdots &  \vdots &  \vdots &  \vdots &  \vdots &  \vdots &  \vdots &\vdots &  \vdots &  \vdots &  \vdots \\
HD\,204827\,(n2)\tablenotemark{c} & ``V'' & 0.550 &  $-$2.83 &  0.14 &   4.88 &  0.14 &  5.63 &  0.14 &   60.1 &  0.7 &  \phn6.60 \\
HD\,204827\,(n2)\tablenotemark{c} &  ``R'' & 0.650 &  $-$2.77 &  0.10 &   4.34 &  0.10 &  5.14 &  0.10 &   61.3 &  0.6 &  12.0\phn \\
HD\,204827\,(n2)\tablenotemark{c} &  ``I'' & 0.800 &  $-$2.48 &  0.16 &   3.61 &  0.16 &  4.38 &  0.16 &   62.2 &  1.1 &  \phn2.40 \\
\vdots &  \vdots &  \vdots &  \vdots &  \vdots &  \vdots &  \vdots
&\vdots &  \vdots &  \vdots &  \vdots \\
\enddata

\tablenotetext{a}{Data are reported in 11 wavelength bins with centers
  spanning 0.485--0.985~$\micron$ with widths of
  $0.050\,\micron$ ("Narrow"). Also shown are data in three bins with centers
  0.550, 0.650, and 0.800~$\micron$ with full-widths of 0.100, 0.100,
  and 0.150 $\micron$, respectively; we refer to these as the $V$-,
  $R$-, and $I$-like broadband filters.}
\tablenotetext{b}{Polarization position angle measured east of north.}
\tablenotetext{c}{Data for HD\,204827 are given for three separate
  nights of observations labeled n2, n3, and n4.}

\tablecomments{All polarization data for the standard
  high-polarization stars used here. Listed
  uncertainties are statistical only and are returned as part of the
  fitting procedure, along with the reduced-$\chi^2$ reported in the
  last column; uncertainties here do not include other systematics
  discussed in the text. (This table is available in its entirety in a
  machine-readable form in the online journal. A portion is shown here
  for guidance regarding its form and content.)}

\end{deluxetable}

The star HD\,212311 was observed as a standard unpolarized star
\citep{turnshek1990} during all three nights of observations.  Due to
the low level of instrument polarization, there was insufficient
signal-to-noise to measure it independently in each wavelength bin.
Averaging the fit Stokes parameters across all wavelength bins yielded
0.1\%--0.3\%, consistent with other calibrations of the Kast
spectropolarimeter (Ryan Chornock 2010, private communication;
\citealt{leonard2001,chornock2010,eswaraiah2012}). The star PPM\,93241
was used as a secondary unpolarized standard based on our more precise
TurPol results for that star
(Table~\ref{tbl-results-turpol}). Measured Kast values are also
0.1\%--0.3\%, compared to the TurPol results that average
$\sim$$0.1\%$--0.2\%.
Since these low levels of instrument polarization are comparable to
the statistical and systematic uncertainties in the Kast data, we made
no correction for the instrument polarization.

To facilitate quantitative comparison with the broadband observations,
we define $V$-,
$R$-,
and $I$-like
passbands for the Kast spectra with centers at 0.55, 0.65, and 0.80
$\micron$,
with widths of 0.10, 0.10, and 0.15 $\micron$,
respectively, and uniform responses across those widths. The results
for the standard stars in these passbands are
given in Table~\ref{tbl-polstds}.  The angle values are in good
agreement, within the statistical uncertainties on the measurements
and the angle calibration (a total of
$\sim$$1.5\arcdeg$--$2\arcdeg$;
Appendix~\ref{sec-angcal}).  With the exception of an outlier in one
measurement of HD\,204827 (night-4, 0.935\,\micron), these stellar
data are all consistent with wavelength-independent position angles.
Absolute deviations of the measured polarization amplitudes from their
``expected'' values are in the 0.0\%--0.4\% range.  We attribute this
level of deviation to contributions from the statistical uncertainties
(0.1\%--0.2\%) and the instrument polarization (0.1\%--0.3\%).

\subsection{Polarization Efficiency}
\label{sec-poleff}

To measure the polarization efficiency of the Kast instrument, a
polarizing filter is inserted just before the HWP \citep{miller1988}.
Standard polarization measurements are then performed using known
standard unpolarized stars to illuminate the slit. Over the course of
three nights we obtained data using four separate observations of
PPM\,93241 and three observations of HD\,212311.  These seven
observations were treated as independent observations and reduced as
described in Appendix~\ref{sec-pfits}.
In the range 0.460--0.860\,$\micron$ the median efficiencies are
99.7\%--99.9\%; the standard deviation across the seven measurements
is $\lesssim$$0.1\%$.
Beyond $0.860\,\micron$
the efficiency drops rapidly, reaching about 50\% by
$1.000\,\micron$.
This sharp drop was not seen in any of the standard high-polarization
stars we observed (Section~\ref{sec-highp}). Therefore, we attribute
this drop not to any down-stream optical element in the instrument,
but to the polarizing filter itself. This is consistent with work by
other users of the Kast polarimeter (Ryan Chornuck, private
communication; \citealt{miller1988}). Given measurements of such high
efficiency, we made no corrections in any of the polarization
measurements made herein.

\subsection{Polarization Angle Calibration} \label{sec-angcal}

The polarization position angle $\theta$ of a celestial source
projected on the sky is related to the measured HWP phase angle
$2\delta$ (equation~[\ref{eq-delta}]), the rotation angle of the
instrument, $\gamma$, and the orientation of the HWP's fast and slow
axes, $\chi$, via the relation
\begin{equation}
  \theta(\lambda) = \gamma - 2\delta(\lambda) + \chi(\lambda) - \beta.
  \label{eq-angles}
\end{equation}
The instrument rotation angle $\gamma$ is the angle of an instrument
axis measured east of north.  The ``zero-angle'' of the HWP,
$\chi(\lambda)$, was measured by inserting a polarizing filter into
the optical path.  This quantity measures the angle with respect to
the linearly polarizing axis of this filter. The angle $\beta$
describes the remaining angle between the instrument axis which
defines $\gamma$ and the axis of the polarizing grid; this was
measured by comparing the measured angles with those of standard
high-polarization stars.


When the polarizing filter was in place, we set
$\theta=\gamma=\beta=0$ in equation (\ref{eq-angles}) yielding
$\chi(\lambda)=2\delta(\lambda)$. This angle varied by nearly
$5\arcdeg$ across the 0.460--1.000~$\micron$ range.  The trend in
wavelength was consistent across different measurements with relative
offsets as large as $\sim$$0.3\arcdeg$.
This offset was consistent with the repeatability of placing either
the HWP or the polarizing filter at any given angular position
\citep{miller1988}. Therefore, we assigned a
$\sim$$0.3\arcdeg$
systematic uncertainty to all angle analyses in this work.

To correct for the wavelength dependence of the phase angle, we used
the median angles that resulted after shifting all angles to agree at
$R$-band; we set this value to zero degrees at $R$-band.  To check
this correction, we examined the residuals after subtracting the
median angles from the shifted observations. The variation across
wavelengths, which was previously as large as $5\arcdeg$, was reduced
to $\lesssim$$0.3\arcdeg$.

In order to measure $\beta$,
we used the angles measured in the synthetic broadband $V$-,
$R$-,
and $I$-like
filters (corrected for the instrument angle $\gamma$
and the wavelength-dependent HWP-zero angle $\chi$)
for BD+25727 and HD\,204827 and compared them to previous measurements
of those same stars.  These three filters and two stars resulted in 12
separate measurements of $\beta$
with a median of $-96.8\arcdeg$
and a standard deviation of $1.3\arcdeg$.
The measured angles in Tables~\ref{tbl-results} and \ref{tbl-polstds}
were corrected using this value. Quadratically combining the angle
uncertainties from the fitted data, uncertanties in HWP-zero angle
($\sim$$0.3\arcdeg$),
and the uncertainty on $\beta$ reported here yielded typical angle
uncertainties $\sim$$1.5\arcdeg$--$2\arcdeg$.
These uncertainty estimates were consistent with the repeatability of
the HD\,204827 angle estimates.  Additionally, the resulting broadband
angles were in good agreement with the TurPol data for PPM\,93776 and
PPM\,93780.

\section{Serkowski \& Wilking Fits} \label{sec-serkfits}

The polarization-wavelength relation in equation~(\ref{serkowski}) can
be written as a polynomial that is linear in its coefficients
\citep[e.g.,][]{coyneetal1974}:
\begin{equation}
  \ln p = X_1+ X_2\ln\lambda - K\ln^2\!\lambda, \label{serk_lin1}
\end{equation}
where
\begin{eqnarray}
  X_1 & = & \ln p_\mathrm{max} - K\,\ln^2\!\lambda_\mathrm{max}, \label{serk_lin2}\\
  X_2 & = & 2K\,\ln\lambda_\mathrm{max}. \label{serk_lin3}
\end{eqnarray}
The solution for the coefficients $X_1$, $X_2$, and $K$ is a linear
regression, similar to that in Section~\ref{sec-pfits}.  The values
$p_\mathrm{max}$ and $\lambda_\mathrm{max}$ are then found from
equations~(\ref{serk_lin2})--(\ref{serk_lin3}).

The reduced-$\chi^2$ values in Tables~\ref{tbl-serkowskifits-turpol}
and \ref{tbl-serkowskifits} utilize the fits to
equation~(\ref{serk_lin1}).  This applies to both the Kast and Turpol
data.

\bibliography{bgbiblio_tot_2}

\end{document}